\documentclass[useAMS,usenatbib]{mn2e}
\usepackage{amssymb}
\usepackage[pdftex]{graphicx}
\usepackage[pdftex,usenames,dvipsnames]{xcolor}
\usepackage{natbib}
\usepackage{aas_macros}
\usepackage{url}
\usepackage[none]{hyphenat}
\usepackage{times}
\usepackage[linktocpage=true,backref=true,pagebackref=false,bookmarks=true,bookmarksnumbered=False,colorlinks=true,linkcolor=NavyBlue,citecolor=NavyBlue,urlcolor=NavyBlue]{hyperref}
\usepackage[all]{hypcap}
\usepackage[backgroundcolor=yellow,linecolor=red,bordercolor=yellow]{todonotes}

\pdfpageattr {/Group << /S /Transparency /I true /CS /DeviceRGB>>} 

\title[Nuclear astrometry in type 2 galaxies]{The central parsecs of active galactic nuclei: challenges to the torus\thanks{Based on European Southern Observatory VLT programs \mbox{076.B-0493}, 084.B-0568 and {\mbox{070.B-0409}}.}}
\author[Prieto et al.]{M.\,A. Prieto$^{1,2}$\thanks{Email: aprieto@iac.es},
M. Mezcua$^{1,2}$, J.\,A. Fern\'andez-Ontiveros$^3$, M. Schartmann$^{4,5}$ \\
 $^1$Instituto de Astrof\'isica de Canarias (IAC), E--38200 La Laguna, Tenerife, Spain \\
 $^2$Universidad de La Laguna, Dept. Astrof\'isica, E--38206 La Laguna, Tenerife, Spain\\
 $^3$Max-Planck-Institut f\"ur Radioastronomie (MPIfR), Auf dem H\"ugel 69, D--53121 Bonn, Germany\\
 $4$ Max-Planck-Institut f\"ur Extraterrestrische Physik, Giessenbachstrasse, D-85748 Garching, Germany\\  
 $5$ Universit\"ats-Sternwarte M\"unchen, Scheinerstrasse 1, D-81679 M\"unchen, Germany}
\date{}

\begin{document}

\label{firstpage}

\maketitle

\begin{abstract}
Type 2 AGN are by definition nuclei in which the broad-line region and continuum light are hidden at optical/UV wavelengths by dust. Via accurate registration of infrared (IR) Very Large Telescope adaptive optics images with optical \textit{Hubble Space Telescope} images we unambiguously identify the precise location of the nucleus of a sample of nearby, type 2 AGN. Dust extinction maps of the central few kpc of these galaxies are constructed from optical-IR colour images, which allow tracing the dust morphology at scales of few pc. 

In almost all cases, the IR nucleus is shifted by several tens of pc from the optical peak and its location is behind a dust filament, prompting to this being a major, if not the only, cause of the nucleus obscuration. These nuclear dust lanes have extinctions $A_V \geq 3-6$ mag, sufficient to at least hide the low-luminosity AGN class, and in some cases are observed to connect with kpc-scale dust structures, suggesting that these are the nuclear fueling channels.

A precise location of the ionised gas H$\alpha$ and [\textsc{Si\,vii}] $2.48\, \rm{\micron}$ coronal emission lines relative to those of the IR nucleus and dust is determined. The H$\alpha$ peak emission is often shifted from the nucleus location and its sometimes conical morphology appears not to be caused by a nuclear --torus-- collimation but to be strictly defined by the morphology of the nuclear dust lanes. Conversely, [\textsc{Si\,vii}] $2.48\, \rm{\micron}$ emission, less subjected to dust extinction, reflects the truly, rather isotropic, distribution of the ionised gas.

All together, the precise location of the dust, ionised gas and nucleus is found compelling enough to cast doubts on the universality of the pc-scale torus and supports its vanishing in low-luminosity AGN. 

Finally, we provide the most accurate position of the NGC 1068 nucleus, located at the South vertex of cloud B.

 \end{abstract}


\section{Introduction}
The centre of galaxies are often very dusty regions. The existence of this dust, either placed at the very core region, a toroidal structure, or being foreground dust lanes crossing the central region or seen in projection, should obscure the nucleus and in some cases hide it completely even up to the infrared (IR) light. Surprisingly enough, the canonical classification of active galactic nuclei (AGN) in type 1/2 sources is done at optical wavelengths, with type 1 --unobscured sources-- having optical broad emission lines and featureless continuum and type 2 --obscured nuclei-- presenting optical narrow-emission lines and continuum dominated by host galaxy light. However, in those type 2 cases where the nucleus is completely obscured, one may wonder on which region the AGN classification is done.\\

The presence of dust in the central region may inevitably lead to misleading identifications of the nuclear source and/or of its nature, prompted by confusion with e.g. a young stellar cluster (this has been the case for almost 20 years in the prototype starburst galaxy NGC\,253; \citealt{2009MNRAS.392L..16F}) or with an ionised gas cloud (this is for example the case in the Seyfert 2 galaxy NGC\,1386, where the nucleus is confused with a bright blob in the extended ionised gas; \citealt{2000ApJS..128..139F}; Fig.\,\ref{NGC1386}, this work). Nuclear dust can also mislead the classification of galaxies as being e.g. non nucleated --a clear example of this is the case of the Seyfert type 2 NGC\,3169, classified as non-nucleated on the basis of \textit{Hubble Space Telescope} (\textit{HST}) data \citep{2008AJ....135..747G} when in reality it shows an outstanding point-like source longwards of $2\, \rm{\micron}$ (see Fig.\,\ref{NGC3169}, this work). The confusion leads to an unrealistic determination of key parameters such as the nuclear luminosity and power, and untrue spectral energy distributions (SED) mixing IR nuclear fluxes with optical off-nuclear measurements. The misidentification propagates directly into other ranges of the spectrum, for example to radio or X-rays, where the core may be associated with the wrong optical or even IR source --this has been also the case in e.g. NGC\,253 \citep{2010ApJ...716.1166M}.

This work shows that IR observations at the pivot wavelength of $2\, \rm{\micron}$, combined with high angular resolution, are key to disclose unambiguously the nucleus of a galaxy. In the course of our studies of high-spatial-resolution images in the IR of the nearest AGN, we found that $2\, \rm{\micron}$ is a key wavelength at which AGN, regardless of their class, show an overwhelming bright point-like source at the centre (by far the brightest source in the central kiloparsec region), naturally classified as the nucleus (\citealt{2010MNRAS.402..724P}). In this paper we focus on all type 2 and intermediate cases --i.e. those with an obscured nucleus-- in our study sample and show that the position of this IR source lies several tens of parsec (pc) away from the optical photometric peak.

The location of the nucleus in optical images is performed via accurate astrometry between optical \textit{HST} and IR Very Large Telescope (VLT) adaptive optics (AO) images in which several point-like sources are identified in the common field of view (FoV) to all images. The later are deep enough for identification of several (often two to three) plausible stellar clusters in the galaxy, seen simultaneously at frequencies in and about the \textit{V}-, \textit{I}- and \textit{K}-bands. Having the image sets comparable angular scales, accurate image registration --in the $10$ to $40\, \rm{mas}$ range-- is performed by relying only on the centroid position of these point-like sources (the nucleus is never used). As a result, the location of the IR peak --the nucleus-- in the optical images is determined. We further present it in IR-optical colour images, in general $I - K$, that effectively illustrate the location and morphology of the central dust distribution. 
Finally, by using the same alignment procedure, we show the distribution of the ionised gas relative to the IR nucleus. The later is expected to be located at the vertex of the ionisation cones and associated with strong coronal line emission (e.g. [\textsc{Si\,vii}] at $2.48\, \rm{\micron}$, with an ionisation potential of $205\, \rm{eV}$).

Image registration based on a similar approach has been reported in the literature for a few objects, which are summarized below. In all cases, the nucleus is identified with the bright central source that dominates the emission in IR images. In the low-luminosity radio galaxy Centaurus A, there is not even an optical photometric peak due to its dense enshrouding dust lane. The nucleus is discovered already in the \textit{I}-band (\citealt{2000ApJ...528..276M}) and the precision of its location in optical images relative to stars in the field is $\sim 30\, \rm{mas}$. In the Seyfert type 2 Circinus galaxy, the optical photometric peak is found $\sim 150\, \rm{mas}$ northwest of the \textit{K}-band peak. The optical photometric peak in this case is a diffuse elongated structure at the vertex of the ionisation gas cone (\citealt{2004ApJ...614..135P}). In the starburst galaxy NGC\,253, the nuclear region includes multitude of stellar clusters and dust. The nucleus is identified with a flat spectrum radio source that has no counterpart yet at any wavelength (including the IR). The location of the kinematic centre, which has no obvious visible counterpart, is found with an error of $\sim 0\farcs5$ relative to the position of the radio core (\citealt{2009MNRAS.392L..16F}; \citealt{2010ApJ...716.1166M}). Finally, a precise determination of the nucleus location in the Seyfert type 2 NGC\,1068 has been extensively sought by many authors, using very different procedures that include accurate coordinates precision, direct astrometry and determination of the UV centre-polarisation pattern (this assumed to be the nucleus; see a further discussion of these works in Appendix~\ref{individual}). In this work we present what in our view is the most accurate location of the nucleus of this galaxy to date, which is independent on any ad-hoc assumption but that the IR bright central source is the nucleus. We find this source to be at $110 \pm 30\, \rm{mas}$ off the optical peak, the location being in agreement with the polarisation centre determined by \citet{1999ApJ...518..676K}.

 
This paper reports on the nucleus location in seven nearby AGN: five Seyfert type 2 (NGC\,1068, NGC\,1386, NGC\,7582, ESO\,428-G14 and MGC-05-23-016), and two low-luminosity AGN (LLAGN) type 2 LINER (NGC\,3169 and NGC\,4594, also known as Sombrero galaxy). All are obscured nuclei in the optical with the exception of the Sombrero galaxy. The latter is included as an example of a true type 2 source, i.e. not obscured by dust at any range of the spectrum --classified so due to the lack of broad emission lines \citep{1998MNRAS.300..893N,2010ApJ...714..115S}. The targets are part of a larger sample that includes the nearest and brightest galactic nuclei that could be observed with AO in the IR from the ESO-Paranal observatory. The detailed analysis of the central pc of these galaxies based on this unprecedented high-spatial-resolution IR data approach is presented in a number of papers: NGC\,1068 and NGC\,7582 in \citet{2010MNRAS.402..724P}; ESO\,428-G14 in \citet{2005MNRAS.364L..28P}; NGC\,1386, NGC\,3169 and Sombrero in \citet[][2014 in preparation]{2012JPhCS.372a2006F}.

A description of the data and analysis is provided in Sections\,\ref{data} and \ref{analysis}, the results are presented and discussed in Section\,\ref{results}, while final conclusions are provided in Section\,\ref{conclusions}. An individual discussion for each object is included in Appendix\,\ref{individual}.

\begin{table*}
\begin{minipage}{\textwidth}
\centering
\caption{Filters, errors and peak distances.}\label{filtros}
\begin{tabular}{lccccccccc}
\hline
\hline 
Object          & $D_\mathrm{L}$ & 1$\arcsec$/pc  &      Ref. & Nucleus  & Nucleus  & Ionised  &    Clusters    &    Errors                          &  Optical-IR \\
	        &      [Mpc]    &        &    & observed  &  obscured  &  gas  &         &[mas]            &    distance [mas]       \\ 
(1)             &       (2)       &      (3)      &            (4)           &          (5)        &    (6)           &         (7)           &         (8)          &       (9)  &  (10)                   \\ 
\hline
ESO\,428-G14	&	19	 & 92 &   [1]      &	        IB2.42	     &	     F814W	   &	H$\alpha$, [\textsc{Si\,vii}]$^{a}$       &	3	         &  40,30,10      &      50 (5 pc)   \\
MCG-05-23-016   &      38  & 180       &   [2]      &        \textit{Ks},F791W &	    F547M          & H$\alpha$$^{b}$  &         2	         &    30,30       &      80 (14 pc)  \\
NGC\,1068	&     14.1       & 70 &   [3]      &         \textit{Ks}	     & F791W   & [\textsc{Si\,vii}]$^{c}$, [\textsc{O\,iii}]          &         3                &  30,20,40         &      110 (8 pc)   \\
NGC\,1386	&      15.3      & 74  &   [4]      &         \textit{Ks}	     & F814W   & H$\alpha$, [\textsc{Si\,vii}]$^{a}$          &         2                &  10,10         &      230 (17 pc)   \\
NGC\,3169	&      24.7     & 120  &   [5]      &            \textit{Ks}   & F814W	   &  H$\alpha$      &         3	         &   20,10,30     &     140 (17 pc)   \\
Sombrero 	&       9.08    & 44  &   [4]      &  F250W--\textit{Ks}      &	       --          & H$\alpha$	      &	        3		 &    10,10,10    &      --    \\
(NGC\,4594)      &                 &            &                          &                     &                                  &                          &                &            \\
NGC\,7582	&      19.9     & 96  &   [6]      &         F160W, IB2.06     &      F606W$^e$	   &  [\textsc{Si\,vii}]	      &         4	         &   20,50,70,20  &      --    \\
  \hline
\end{tabular}
\end{minipage}
\raggedright
\smallskip\newline\small {\bf Column designation:}~(1) Object name; (2) luminosity distance; (3) linear scale; (4) references for column (2): [1] \cite{1996ApJ...470L..31F}; [2] \cite{2000ApJS..128..139F}; [3] \cite{1997A&A...320..399M}; [4] \cite{2003ApJ...583..712J}; [5] \cite{2009ApJ...704..629M}; [6] \cite{2002A&A...393...57T}; (5) bands in which the nucleus is observed; (6) reddest band in which the nucleus is obscured; (7) ionised gas; (8) number of stars/clusters used in the alignment; (9) error on the position of each star/cluster, averaged over all the filters mentioned in columns 5--7; (10) distance between the optical (bluest band)- and the IR-peak emission.\\
$^a$ for ESO\,428-G14 and NGC\,1386, the [\textsc{Si\,vii}] image has been aligned using the IB2.42 nucleus; $^{b}$ for MCG-05-23-016, the H$\alpha$+[\textsc{N\,ii}] image has been aligned using only 1 star; $^{c}$ for NGC\,1068, the [\textsc{Si\,vii}] image has been aligned using the \textit{Ks} nucleus; $^e$ in NGC\,7582, faint emission at the location of the IR nucleus is seen in this filter, variable within a factor of $\sim 1.5$ between 1995 and 2001.
\end{table*}

\section{Source of data}
\label{data}
\subsection*{Adaptive optics IR data}
The sample of type 2 AGN was observed with the ESO AO-assisted IR camera NaCo (Nasmyth Adaptive Optics System --NAOS-- plus Near-Infrared Imager and Spectrograph --CONICA) at the VLT in the $1$--$5\, \rm{\micron}$ range. In all cases, the nucleus was used as reference for the AO correction, which is the optimal method to maximize the attainable spatial resolution. The NaCo FoV covers the central few kpc of each galaxy. In this paper, we make use of the broad and narrow band images taken in the $2\, \rm{\micron}$ band. We selected the $2\, \rm{\micron}$ images as this is the pivotal band where the contrast between the AGN and the host galaxy reaches its maximum (\citealt{2010MNRAS.402..724P} and references therein). Specifically, the filters used are the broad-band \textit{Ks} ($\lambda_c = 2.180\, \rm{\micron}$, $\Delta\lambda = 0.350\, \rm{\micron}$) image and the two intermediate-band line-free filters IB2.42 ($\lambda_c = 2.420\, \rm{\micron}$, $\Delta\lambda = 0.060\, \rm{\micron}$) and IB2.06 ($\lambda_c = 2.060\, \rm{\micron}$, $\Delta\lambda = 0.060\, \rm{\micron}$). We also made use of our \textit{J}- and \textit{H}-bands images --for confirmation of the nucleus location-- although these are not explicitly presented here. In addition, the intermediate band IB2.48 ($\lambda_c = 2.480\, \rm{\micron}$, $\Delta\lambda = 0.060\, \rm{\micron}$) centred on the high ionisation [\textsc{Si\,vii}] $2.48\, \rm{\micron}$ coronal emission line is also used in this analysis, after subtraction of the adjacent-continuum (IB2.42 band). These line data serve as a tracer of the central ionising source location.

The NaCo data reduction was performed using the ESO \textsc{eclipse} package \citep{1999ASPC..172..333D}. The reduction algorithm includes sky subtraction, registration and combination of individual frames for each dataset.

\subsection*{\textit{HST} optical data}
The widest possible spectral range from UV to \textit{I}-band in \textit{HST} images was surveyed in a quest to identify the reddest wavelength at which the nucleus, if any, starts to show up and to identify the UV-optical peaks. Table\,\ref{filtros} lists the filters examined for each galaxy. For the alignment procedure we used the reddest available image, which in general provides the larger number of reference sources common to the $2\, \rm{\micron}$ NaCo images.

For those galaxies without [\textsc{Si\,vii}] images available --three cases, Table\,\ref{filtros}-- H$\alpha$+[\textsc{N\,ii}] images were constructed from corresponding \textit{HST} images centred on the blend and on the broad-band continuum (typically \textit{V} or \textit{I}-bands), and using customary procedures for pure-line emission extraction. When available, the continuum was derived from the interpolation between \textit{V} and \textit{I}-bands, i.e. this is the case of NGC\,1386, NGC\,3169 and Sombrero.

In order to achieve a high precision in the determination of the relative shift between the IR and optical peaks, the \textit{HST} archival retrieved images were subjected to a more accurate geometric distortion correction. This was done with the \textit{MultiDrizzle} package in PyRAF (\citealt{2003hstc.conf..337K}). For datasets with single exposures, cosmic rays removal was done using the L.A. Cosmic algorithm \citep{2001PASP..113.1420V} in its \textit{Python} module version\footnote{\url{http://python.org}\\ \url{http://obswww.unige.ch/~tewes/cosmics_dot_py/}}. Further analysis was done with \textsc{iraf}/PyRAF.

\section{Analysis}
\label{analysis}
\subsection{Image registration: nucleus location}
Figs\,\ref{ESO428}--\ref{NGC7582} show, for each galaxy, all the registered images that are used. In all cases, the top left panel shows an optical \textit{HST} image (typically \textit{V}- or \textit{I}-band), 
with NaCo $2\, \rm{\micron}$ contours overplotted (either \textit{Ks}-band --most cases-- or the narrow-band filters centred on pure continuum IB2.42 and IB2.06, see Table\,\ref{filtros}). The point-like sources common to the FoV to all optical and IR images are marked with blue circles. These are usually stellar clusters in the galaxy body or, exceptionally, foreground stars. Overall, up to two to three point-like sources in the common FoV of the images could be identified, the major limitation being both the reduced FoV and higher galaxy emission in the near-IR.

Image registration was done with the IRAF task \textit{imalign} and crosschecked with alternative image alignment programs written by us in IDL\footnote{Interactive Data Language.}. For each galaxy, we use \textit{HST} images centred at wavelengths around $3000$, $5000$ and $8000\, \rm{\AA}$ to register with NaCo $2\, \rm{\micron}$ images. In the case of the galaxies ESO\,428-G14, NGC\,3169 and Sombrero, the image registration required the orientation of the NaCo \textit{K}-band image to be rotated by a small angle of the order of a few tenths of degree, which corresponds to the precision of the rotators in the instrument \citep{2008JPhCS.131a2028K}. Relative offsets between images were determined from the reference point-like sources identified in the FoV, the $2\, \rm{\micron}$ nucleus was never used as a reference source to align optical images. The IR nucleus was only used to align the IB2.48 filter with the IB2.42 filter and the \textit{Ks}-band in the cases of ESO\,428-G14 and NGC\,1386, respectively, i.e. long-wards of $2\, \rm{\micron}$ where it is an outstanding source.

Images were shifted using the average from those offsets (listed in Table\,\ref{filtros}). For each reference source, we provide an alignment error estimated as the standard deviation of its position among the different optical and IR filters used. 
The errors of the reference sources for each galaxy do not diverge much but are consistent within 20--30 mas. The reference sources are, for most galaxies, either randomly distributed or in triangulation. Thus, these errors account for uncorrected distortions, wavelength-dependent positions and gradients in the background emission that affect the estimate of the centroid position. The precision of the alignment is taken as the mean of the individual errors of all the reference point-like sources used. Therefore, the final errors implicitly include the residuals of geometrical distortions. These final alignment errors are on average between $10$ and $40\, \rm{mas}$ (see Table\,\ref{filtros}), which correspond to $\sim$0.2--0.8 pixels (see Table\,\ref{resolution}). Our astrometry is thus better than the pixel size in all sources.

\begin{table}
\centering
\caption{Highest pixel scale and filter}
\label{resolution}
\begin{tabular}{lcc}
\hline
\hline 
Object	  &	1$\arcsec$/px & Filter\\
\hline
ESO\,428-G14	&  0.0455	& IB2.42\\                       
MCG-05-23-016 	&  0.1          & F547M\\
NGC\,1068	&  0.0455	& F791W\\
NGC\,1386	&  0.0455	& F814W\\
NGC\,3169	&  0.05	        & F814W\\
Sombrero (NGC\,4594) &  0.05	& F435W\\
NGC\,7582	&  0.075	& F160W\\
\hline
\end{tabular}
\end{table}

\subsection{Dust extinction maps: the nucleus hidden by dust filaments}
\label{dust}
The optical-IR aligned images were used to construct colour maps for the central few kpc region of each galaxy (Figs\,\ref{ESO428}--\ref{NGC7582}). These maps effectively trace the dust distribution in the central region. The colour maps shown in the figures are in most cases a ratio between the NaCo $2\, \rm{\micron}$ and \textit{HST} optical continuum images, preferentially at \textit{V}- or \textit{I}-bands. All the maps are characterized by a relatively-contrasted point-like source at the centre of the image that coincides with the location of the $2\, \rm{\micron}$ peak. In moderately luminous nuclei, particularly in the LLAGN class (including Sombrero and NGC\,3169), the colour maps show relatively faint emission at the nucleus $2\, \rm{\micron}$ peak location due to the low contrast between the galaxy and the nucleus. The colour maps also illustrate the ubiquitous presence of dust in all the galaxies within the central, about hundred pc, region. For most sources, this central dust is seen associated with a much global dust structure extending over kpc scale. The dust has a varied morphology --as projected on the sky-- from filamentary or disk-like to more amorphous, undefined shapes.

An estimate of the extinction $A_V$ introduced by this central dust structure in the surroundings of the nucleus was inferred by comparing galaxy colours at dusty regions near to the centre, e.g. along a filament crossing the centre, with those at further away locations showing no apparent signatures of dust extinction (these would correspond to darker regions in the colour maps; dusty regions are shown as bright). Overall, colours of dust dominated regions were measured at distances of between 20 and 90 pc from the centre depending on the object; those of dust-free at distances $\lesssim 0.5\, \rm{kpc}$ (see Table~\ref{extinction}). The values reported in the table correspond to the average of three to four values measured at the indicated distances from the centre. The size of the aperture radius used to measure the colours in the dusty regions was adapted to the width of the observed filaments, disks or lanes.
For the dust-free regions, typically aperture radius of 0.1--0.2$\arcsec$ are used in order to obtain a reliable estimate of the unreddened colour of the galaxy.
These reference regions, supposed to directly trace the underlying stellar galaxy population, may still be subjected to some dust extinction, although at lower level. Thus, the inferred $A_V$ values may be lower limits. For consistency check, $A_V$ values were determined --when available-- from two different colour maps we could produce from the set of registered images (see Table\,\ref{extinction}). Maps involving the widest wavelength range are optimal to enhance the dust location and morphology, yet they could sometimes lead to misleading results if the optical broad-band filters include important contribution from lines as [\textsc{O\,iii}] or H$\alpha$. 
In producing the $A_V$ values listed in Table\,\ref{extinction}; the extinction law from \citet{1999PASP..111...63F} applies.


\begin{table*}
\centering
\label{extinction}
\begin{tabular}{lcccccccccc}
\hline
\hline 
          & Galaxy &  AGN	&  &		&  $A_V$ mag&          &     &   Visible nucleus &  \multicolumn{2}{c}{Nuclear distance} \\
Object	  &	Type & Type & $U - K$ &	$V - K$ & $V - I$ & $R - K$ & $I - K$ & \textit{HST} filter & Dust & Dust-free \\
(1)		&   (2)     &     (3)    &   (4)     &    (5)    & (6)   &   (7)   & (8)   & (9)    &   (10)   &  (11) \\    
\hline
ESO\,428-G14&S0 & LLAGN-2  &  -& -    &- & -  & 3& no & 0.4$\arcsec$ & 2.5$\arcsec$ \\                       
MCG-05-23-016&S0a & Sy-2  &  -& 1.3& 0.9 & - & - &  F814W & 0.6$\arcsec$ & 2$\arcsec$--4$\arcsec$\\
NGC\,1068&Sb   & Sy-2  & - & 2.4 & -& -& 6& no & 0.4$\arcsec$--0.5$\arcsec$ & 7$\arcsec$\\
NGC\,1386&S0a   & LLAGN-2  & -& - & - & - & 3 & no & 0.5$\arcsec$ & 4.5$\arcsec$\\
NGC\,3169&Sa  & LLAGN-2  &  -&  2.5& -& - & 2.7&  no & 0.3$\arcsec$--0.5$\arcsec$ & 6.5$\arcsec$\\
Sombrero&Sa  & LLAGN-2  & 0.5& -& -& -& 0.8& F250W & 0.4$\arcsec$ & 1.8$\arcsec$\\
(NGC\,4594)  &   &  &  &  &    &   &   &      &          \\
NGC\,7582&SBab   & Sy-2  & -& -& - & 5.2& -& F606W & 0.3$\arcsec$  & 3.5$\arcsec$\\ 
\hline
\end{tabular}
\caption{Extinction estimate derived from colour maps. For consistency check, when possible two different $A_{V}$ are estimated from best suitable continuum images. Dust-dominated and dust-free regions were selected at the radius from the centre indicated in the last two columns, those of dust-free correspond to $\lesssim 0.5\, \rm{kpc}$. Several values were measured at these locations and their average is provided in the table. For simplicity, colours are referred to common Johnson system although they are measured in specific \textit{HST} and VLT-NaCo filters as indicated in the text and Table\,\ref{filtros}. Column 9 indicates the optical band in which the nucleus is first visible; "no" means nucleus undetected up to $2\, \rm{\micron}$. The AGN type Sy stands for Seyfert. LLAGN-2 refers to nuclei with $L_\mathrm{bol} < 3\times10^{42}$ erg s$^{-1}$ as integrated from our high-spatial-resolution SEDs \citep{2012JPhCS.372a2006F,2014MNRAS.tmp..315A}. For those sources with higher $L_\mathrm{bol}$ \citep{2010MNRAS.402..724P,2010MNRAS.402..879R}, the Sy-2 classification is kept.}
\end{table*}

\subsection{Ionised gas peak emission and the nucleus location}\label{gas_nucl}
The location of the ionised gas peak emission is often used as a tracer of the central source location. This is so because ionised gas in AGN often distributes along a preferential direction, in some cases defining a conical morphology whose vertex points or falls onto the nucleus. Optical [\textsc{O\,iii}] and H$\alpha$ lines are the most common lines to trace this emission, and the peak of the emission or the vertex of these conical structures, if not masked by dust extinction, should point to the nucleus location. For this purpose we used the \textit{HST} H$\alpha$ and, when available, the NaCo [\textsc{Si\,vii}] $2.48\, \rm{\micron}$ coronal-line maps. Coronal lines in particular, due to their required ionisation potential, above $100\, \rm{eV}$, are unequivocal tracers of AGN activity. IR coronal lines are in addition excellent tracers because of reduced extinction. The \textit{K}-band [\textsc{Si\,vi}] $1.96\, \rm{\micron}$ and [\textsc{Si\,vii}] $2.48\, \rm{\micron}$ coronal emission lines are among the strongest ones in the IR spectra of Seyfert galaxies (\citealt{2002MNRAS.331..154R,2003MNRAS.343..192R}; \citealt{2005MNRAS.364L..28P}; \citealt{2011ApJ...743..100R}). The later is the strongest, and for half of the galaxies in this work an image registration with high-spatial resolution images in that line was done.
 
The extraction of continuum-free line images follows standard procedures. In the case of the \textit{HST} H$\alpha$ line-plus-continuum images (effectively H$\alpha$ + [\textsc{N\,ii}]), these were registered together with the set of broad-band images combined for each galaxy. The same set of reference point-like sources could be used in all the images (see Table\,\ref{filtros}), thus, the whole image set, line and continuum, are consistently registered to each other. In the case of NaCo coronal line [\textsc{Si\,vii}] and adjacent-continuum images, the global registration followed a different procedure due to the lack of point-like sources (other than the nucleus) in the NaCo narrow-band filters. In this case, the nucleus was used as reference source to align the NaCo narrow-band images with the \textit{K}-band, which was previously aligned to \textit{HST} images. This procedure should not introduce any bias as the nucleus is seen in all galaxies in the $2\, \rm{\micron}$ band. The derived offsets were applied to the [\textsc{Si\,vii}] and adjacent-continuum images, so that the whole dataset (\textit{HST} and NaCo images) become thus registered to a common reference frame. This NaCo procedure was used for ESO\,428-G14, NGC\,1386 and NGC\,7582. For MCG-05-23-016 and NGC 1068, the registration was done with the nucleus and a star/cluster in the common FoV (Table\,\ref{filtros})

Once broad and line images were registered, the adjacent-continuum images were scaled by a factor and subtracted from the line+continuum images to extract pure-line images. This factor was changed iteratively until the emission-line maps did not show negative residuals. The resulting line images are shown in contours in Figs\,\ref{ESO428}--\ref{NGC7582}, bottom panels, on top of the dust extinction colour maps. This comparison illustrates the location of the ionised gas with respect to both the nucleus and the dust.

\section{Results and discussion}\label{results}
The optical image presented in Figs\,\ref{ESO428}--\ref{NGC7582} corresponds to the reddest optical wavelength that was examined and at which the nucleus remains obscured, usually \textit{HST}/\textit{I}-band (see Table\,\ref{filtros}). There are two exceptions: the Sombrero galaxy with the nucleus seen at all optical wavelengths including the UV (Fig.\,\ref{NGC4594}), and MCG-05-23-16 where the nucleus is seen in the \textit{I}-band (Fig.\,\ref{MCG}). NGC\,7582 may be a third exception, it shows faint emission at $6000\, \rm{\AA}$ (\textit{HST}/F606W, Fig.\,\ref{NGC7582}) at the precise location of the \textit{K}-band peak. This emission may relate to the nucleus or to one of the young stellar clusters in the region. We still interpret this counterpart as the nucleus because of three facts: the strength of the emission agrees within the overall SED of the nuclear source \citep{2010MNRAS.402..724P}; the peak of the coronal emission [\textsc{Si\,vii}] $2.48\, \rm{\micron}$ line falls just on that position (see Section\,\ref{ionisedgas}); and the variability shown by the optical counterpart, with a flux increase of a factor of $\sim 1.5$ between 1995 and 2001.

The common characteristic to all the galaxies is the outstanding point-like source at the centre of the NaCo \textit{K}-band images, by far the brightest source in the central few kpc, and thus assumed to be the galaxy nucleus. SEDs compiled for most of these nuclei using very high-spatial-resolution data confirm that indeed the nucleus reaches its maximum luminosity from $2\, \rm{\micron}$ on \citep{2010MNRAS.402..724P,2012JPhCS.372a2006F}. Focusing on the ionised gas, traced by either the [\textsc{Si\,vii}] $2.48\, \rm{\micron}$ coronal line whenever possible and/or H$\alpha$, the peak emission falls in all cases on top of the NaCo \textit{K}-band peak, if the [\textsc{Si\,vii}] line is used, or its vertex points to the \textit{K}-band peak if the H$\alpha$ line is used instead (see bottom panels in Figs\,\ref{ESO428}--\ref{NGC7582}). In the two less extincted nuclei, Sombrero and MCG-05-23-016, the H$\alpha$ peak also falls on top the NaCo \textit{K}-band peak. All together further confirms the IR peak to be the location of the active nucleus.

\subsection {What causes the obscuration of the nucleus?}
The optical-IR colour maps (Figs\,\ref{ESO428}--\ref{NGC7582}, top right) clearly reveal the dust distribution and morphology from the central pc to the kpc scales. The intricate dust morphology seen in these maps contrasts with the uniform, well behaved galaxy light profile traced at $2\, \rm{\micron}$ (white contours in Figs\,\ref{ESO428}--\ref{NGC7582}), which supports the reliability of this image as an adequate tracer of the intrinsic galaxy light. Dust is seen in the centre of all galaxies, up to scales of a few tens of pc from the centre --the spatial resolution of our images. The nucleus appears as a relatively high-contrasted source at the centre of the field, in all cases surrounded, but also crossed, by a complex structure of filaments and lanes, presumably of dust, that extend on the hundred pc scales: e.g. $\sim 1\arcsec$ in ESO\,428-G14 ($92\, \rm{pc}$) and in NGC\,1068 ($70\, \rm{pc}$), $\sim 2\arcsec$ in NGC\,1386 ($148\, \rm{pc}$) and MCG-05-23-016 ($360\, \rm{pc}$), $\sim 4\arcsec$ in Sombrero ($176\, \rm{pc}$) and $\sim 7\arcsec$ in NGC\,7582 ($672\, \rm{pc}$). 

The morphology of the dust filaments is particularly sharp in three remarkable cases: ESO\,428-G014 (Fig.\,\ref{ESO428}, top right) shows a spiraling structure with up to three arms encircling the centre and reaching the nucleus. The spiraling structure is detected further out up to about $1\, \rm{kpc}$ from the nucleus, and it possibly connects with the wider dust lanes that reach the limits of the galaxy body as seen in the \textit{HST} optical image. This nuclear spiral structure is reminiscent of the one seen in the centre of NGC\,1097 \citep{2005AJ....130.1472P}. In NGC\,1386, (Fig.\,\ref{NGC1386}, top right panel) a highly contrasted single dust filament is seen crossing straight the centre. This filament is actually part of a twisted ring which in shape resembles the infinite symbol. Excitingly, it shows an extremely similar appearance to an equivalent one seen at the Galactic Centre, also extending over $100\, \rm{pc}$ scale and crossing the supermassive black hole Sgr A$^*$ \citep{2011ApJ...735L..33M}. In MCG-05-23-016 (Fig.\,\ref{MCG}, top right panel), an edge-on disk, or a single filament, is seen in projection crossing very close to the centre: the nucleus shows just above this filament in the figure. In all other cases, the morphology is undefined but the general trend is for dust always sitting at the centre. This central dust is in most cases seen forming part of a much larger kpc-scale structure (this can be visualized in large-scale-field images in Figs\,\ref{ESO428}--\ref{NGC7582}).

The ubiquitous presence of a central 100-pc-scale dust structure and its strategic location relative to the nucleus naturally argues for it being a major cause, if not the cause, of the nucleus obscuration in optical images. The inferred $A_V$ values towards the centre are moderate, in the $2.5$--$6\, \rm{mag}$ range (Table\,\ref{extinction}). When possible, extinctions were determined from several colour images involving a different short-wavelength image and values tend to agree. Yet, as argued in Section~\ref{dust}, they may still be lower limits. The major uncertainty in determining this extinction is the choice of the dust-free region in the nuclear surrounding of the galaxies. These dust-free regions were selected away from filamentary/lanes structures, in general within a kpc radius from the centre to minimized metallicity or stellar populations changes. An overall impression of the range of $A_V$ values caused by these structures within the central half to 1 kpc region in each galaxy is given in Figs~\ref{extinction1} and \ref{extinction2}.


In some specific cases, the $A_V$ values are most probably lower limits because of the sharp drop-off in H$\alpha$ at locations where these filaments are placed (see Figs\,\ref{ESO428}--\ref{NGC7582}, bottom panels, and next section). Another effect in the $A_V$ determination is the contribution of strong emission lines in the optical images, which may render bluest optical-IR colours, hence lower $A_V$. In the case of NGC\,1068, the colour map shows a complex structure and the extinction inferred from the F550M--\textit{K}-band colour is more than a factor two lower when compared to the F791W--\textit{K}-band colour. This suggests that the optical is likely contaminated by either [\textsc{O\,iii}] or scattered light from the AGN (see Fig.\,\ref{NGC1068_OIII}; Appendix\,\ref{individual}). The highest extinction values are found in NGC\,1068 and NGC\,7582, with $A_V$ above $5\, \rm{mag}$, which can lead to complete obscuration of the nucleus. Intermediate extinctions, $A_V \sim 3\, \rm{mag}$, are found for ESO\,428-G14, NGC\,1386 and NGC\,3169. However, these nuclei are low-luminosity sources, $L_{\rm{bol}} \lesssim 5 \times 10^{42}\, \rm{erg\,s^{-1}}$ (Table\,\ref{extinction}), and thus an extinction of this order can smear out their nuclear emission in the background galaxy light. The lowest extinction is found in Sombrero, $A_V \sim 0.8\, \rm{mag}$, consistent with this being a visible nucleus at all wavelengths including the UV, and in MCG-05-23-016, $A_V \sim 1\, \rm{mag}$, consistent with its nucleus being detected in the \textit{HST}/F791W filter and being presumably partially obscured in the \textit{HST}/F547M filter. In the latter there is a relatively bright source very close to the \textit{K}-band peak position, the shift between both peaks being within the errors (Table\,\ref{filtros}).

%

Putting altogether, the estimated extinctions and the strategic location of the dust in front of the nucleus are sufficient to obscure these nuclei at optical wavelengths and justify the type 2 classification. When $A_V$ is found very low, the nucleus is consistently found fully or partially visible. This finding questions whether a further obscuring structure --the torus-- is needed to explain the lack of broad-line emission and the morphology of the ionised gas. In those cases where well-defined nuclear dust filaments are seen, the filament sizes exceed the $100\, \rm{pc}$ range and in some cases they appear connected to larger kpc-scale dust structures, very much suggesting the flow of material from the outer parts in the galaxy to the very centre, the fueling of the nucleus. By losing angular momentum, this material may naturally lead to a high-optical thickness structure at the centre, perhaps the envisaged pc-scale disk-torus, but a priori, for the sole purpose of hiding the nucleus, the much larger-scale size filaments and lanes shown here could do the obscuration.

\subsection{Relative location of the ionised gas, dust, and the nucleus} \label{ionisedgas}
The location and origin of the ionised gas, in particular of coronal gas, should unequivocally point out the AGN location. The performed astrometry allows us to determine with an accuracy of a few tens of mas (Table\,\ref{filtros}) the position where the ionised gas shows its highest collimation zone. The alignment with NaCo/[\textsc{Si\,vii}] $2.48\, \rm{\micron}$ coronal line, whose origin is undoubtedly the AGN, was possible in four out of the seven galaxies for which that image is available: ESO\,428-G14, NGC\,1068, NGC\,1386, and NGC\,7582. In all other cases, a \textit{HST}/H$\alpha$ line image was used instead.

When the tracer is [\textsc{Si\,vii}] $2.48\, \rm{\micron}$, a strong point-like source is always revealed, in contrast with H$\alpha$ (Figs\,\ref{ESO428}, \ref{NGC1068}, \ref{NGC1386} bottom right panel, and Fig.\,\ref{NGC7582} bottom left panel). The point-like source in [\textsc{Si\,vii}] is thus associated with the position of the nucleus, i.e. the NaCo IR peak (see Section\,\ref{gas_nucl}). In all cases, the extended emission is rather collimated -elongated- at both sides of the nucleus, probably due to the low extinction affecting this line. The same applies for H$\alpha$ but only for those galaxies for which we estimated a very low extinction at the centre, i.e. $A_V$ $\lesssim 1\, \rm{mag}$ in MCG-05-23-016 and Sombrero (Figs\,\ref{MCG} and \ref{NGC4594}, bottom panel). In NGC\,3169 and NGC\,1386, where $A_V$ is $\sim 3\, \rm{mag}$, the peak emission in H$\alpha$ is shifted with respect to the nucleus location due to the extinction suffered by this line: in NGC\,3169 (Fig.\,\ref{NGC3169}, bottom), the H$\alpha$ peak is displaced with respect to the nucleus position by $140 \pm 20\, \rm{mas}$ --the same value as that found with the optical continuum peak (see Table\,\ref{filtros}). In NGC\,1386 (Fig.\,\ref{NGC1386}, bottom left panel), the shift is almost $0\farcs5$ (see next Section). In NGC\,1068 (Fig.\,\ref{NGC1068_OIII}), the shift between the [\textsc{O\,iii}] $5007\, \rm{\AA}$ peak emission and the \textit{K}-band nucleus is however moderate, $50 \pm 30\, \rm{mas}$.

The morphology of the [\textsc{Si\,vii}] $2.48\, \rm{\micron}$ line is not affected by dust extinction in contrast to H$\alpha$. This looks patched in almost all galaxies due to the presence of the dust filaments. For example, in NGC\,3169, H$\alpha$ emission is preferentially at the Western side of the nucleus where there is not much dust, close to the nucleus it extends along the border line defined by the central arc-like dust lane, so the gas shows as well the same arc-like profile at the border line (Fig.\,\ref{NGC3169}, bottom panel). The H$\alpha$ emission disappears at the opposite side due to a thick dust lane and reappears at further distances from the centre at locations where the dust becomes thinner. In NGC\,1386, a sharp cut-off in the ionised gas stretches along the edges of two long dust filaments crossing North-South (Fig.\,\ref{NGC1386}, bottom left panel). No H$\alpha$ is seen on the Eastern side of the galaxy, but most spreads over the apparent dust-free area on the Western side.

The ionised gas of NGC\,7582, in this case traced by a ground-based [\textsc{O\,iii}] $5007\, \rm{\AA}$ image \citep[Fig.\,1 in][]{2009MNRAS.393..783R}, disappears East and North of the nucleus. The fall-off of the emission is outlined by the edges of the dust lanes, this can be seen when comparing Fig.\,1 in \cite{2009MNRAS.393..783R} with our extinction map in Fig.\,\ref{NGC7582} (middle right panel). In ESO\,428-G14 (Fig.\,\ref{ESO428}, bottom left panel), the central spiraling dust filaments somewhat limit the H$\alpha$ emission boundaries. Finally, NGC\,1068 (Fig.\,\ref{NGC1068_OIII}), although with a more complex morphology, shows rather extended [OIII] emission in all directions around the nucleus, yet the outer boundary of the emission is defined by the collar-like bright region seen in the optical-IR color maps. So presumably this is a dusty obscuring region. However, [\textsc{Si\,vii}] is less constrained by this collar-like region
(see also individual discussion in Appendix\,\ref{individual}). Conversely, in the Sombrero galaxy and in MCG-05-23-016, H$\alpha$ does not appear distorted by extinction, which is consistent with the lower values of $A_V$ (below or about $1\, \rm{mag}$). In particular, the dust and gas in Sombrero seem to follow each other (Fig.\,\ref{NGC4594}, bottom) along what presumably is a thin disc in the plane of the galaxy.

 AGN ionisation cones are usually traced by optical gas and their apparently collimated morphology has been widely assumed to be caused by a torus. Amid projection effects, which may cause among other things the asymmetric distribution of the dust seen in some of the galaxies due to inclination, an anti-correlation in position between the ionised gas and the dust is apparent in all sources, from the small to the larger size scales. The observed gas morphology does not require a central nuclear collimation, on the contrary, the location of the dust on larger scales defines the regions where the ionised gas is visible.

If we focus instead on the IR [\textsc{Si\,vii}] line morphology for those objects with this information available, we observe that the emission, which is subjected to much lower extinction, shows accordingly a more isotropic morphology (e.g. a clear example of this is NGC\,7582, Fig.\,\ref{NGC7582}, bottom panel; or the Circinus galaxy, see Fig.\,1 in \citealt{2005MNRAS.364L..28P}, Fig.\,9 in \citealt{2011ApJ...739...69M}), or that it is elongated along the major axis of the galaxy, presumably within a disc in the plane of the galaxy (e.g. ESO\,428-G14 and NGC\,1386, Figs\,\ref{ESO428} and \ref{NGC1386}, bottom right).
 

%

\subsection{Nature of the optical peak emission}
We further examine the position and nature of the optical photometric peak, as this is supposed to be the location where the AGN classification is done and is usually assumed to be the counterpart of the IR nucleus. The Sombrero galaxy is the only case where the photometric peaks at all wavelengths coincide with each other at the level of our astrometry ($10\, \rm{mas}$, $\sim 0.5\, \rm{pc}$) and is thus supposed to be the nucleus. In NGC\,7582, the NaCo $2\, \rm{\micron}$ peak coincides within a precision of $\sim 50\, \rm{mas}$ ($5\, \rm{pc}$) with a faint source in the \textit{HST} optical image. This counterpart resembles a young stellar cluster from the central star-forming region of the galaxy (see Fig.\,\ref{NGC7582}). However, the coronal [\textsc{Si\,vii}] line peak falling on that position (Fig.\,\ref{NGC7582} bottom panel) and the variability detected (a factor of $\sim 1.5$ between 1995 and 2001) strongly suggest that this counterpart must be the active nucleus.

The optical photometric peak in all the other galaxies is shifted from the nucleus by several tens of pc, ranging from $\sim 4.6\, \rm{pc}$ in ESO\,428-G14, to $\sim 7.7\, \rm{pc}$ in NGC\,1068 and up to $\sim 17\, \rm{pc}$ in NGC\,1386, 
well above the astrometry errors (Table\,\ref{filtros}). An equivalent astrometry analysis to the one presented here was carried out in the Seyfert 2 Circinus galaxy and yielded also an optical-peak shift from the IR nucleus of $2.8 \pm 0.4\, \rm{pc}$ \citep{2004ApJ...614..135P}.

In the case of NGC\,1068, several attempts to position its nucleus are reported in the literature. The positioning provided in this work is, as far as we know, the most accurate to date as, contrarily to previous studies, it is independent of any priory assumptions on where the nucleus should be located and/or on the coordinate systems but relies only on the relative position of three point-like sources common to the FoV of all the optical and IR images used. The procedure places the optical peak at $7.7 \pm 1.4\, \rm{pc}$ North from the NaCo $2\, \rm{\micron}$ peak. The optical peak is also the location of the UV peak emission (see Appendix\,\ref{individual} for details) which is often identified as cloud B after \citet{1991ApJ...369L..27E}. Cloud B is interpreted as off-scattered light from the nucleus, and this interpretation is consistent with the emission being slighty extended and the IR nucleus just located at its South vertex (see Fig.\,\ref{NGC1068}, bottom left panel). This morphology arrangement is reminiscent of what is also seen in the Circinus galaxy, where the IR nucleus falls at the vertex of a narrow optical light beam extending over a few pc (see fig.\,2 in \citealt{2004ApJ...614..135P}).

Considering the relatively large optical--IR peak shifts found in the galaxies here studied, it is interesting to question the nature of the optical peak. As we have seen, the optical peak is related to either (1) a cloud in the extended narrow-line region, with the continuum light arising from the host galaxy (good examples are NGC\,3169 or NGC\,1386, in the latter, the brightest H$\alpha$ cloud South of the nucleus has so far been mistaken with the nucleus, e.g. \citealt{2000ApJS..128..139F}); (2) nuclear scattered light in the ionising cone (e.g. Circinus, \citealt{2004ApJ...614..135P}, Fig.\,2); or (3) a young star cluster (e.g. NGC\,7582, this work; NGC\,253 in \citealt{2009MNRAS.392L..16F}). It also further follows that the classical emission line diagnostic diagrams (i.e. BPT diagrams, \citealt{1981PASP...93....5B}) used to classify galaxy nuclei in terms of their activity level are, in the case of obscured nuclei, most probably classifying nebular gas that may be located at different distances from the ionising source, being this an AGN or a star-forming region, or perhaps reflect local ionised gas from an off-nuclear star-forming region.

By definition type 2 nuclei are obscured sources, and the examples of prototypes objects presented in this work show that to be the case as expected. It is thus important to keep in mind that physical parameters such as e.g. AGN luminosities or SEDs derived from the location of the optical peak may not be representative of the true nucleus.

Finally, the results from this work suggest that estimates of the AGN obscured fraction might be mislead by non-torus obscuration. This effect is expected to be particularly strong in those nuclei found in dusty spirals in the intermediate- or low- luminosity range. In this case, a mild obscuration in the optical of a few magnitudes caused by a dust lane is able to hide the active nucleus and mimic a type 2 from a classical pole-on oriented type 1 nucleus. Our sample is limited and does not allow us to quantify this effect, but some individual sources as NGC\,7582 probe this scenario: the AGN in NGC\,7582 is detected in the optical and found to be variable, suggesting that the torus --if present-- is not blocking the line of sight to the central engine. However, the obscuration caused by the dust lane is enough to weaken the strong IR nucleus and turn it into a very faint source, almost indistinguishable from the background light of the host. Obscuring fraction estimates determine current models of the cosmic X-ray background, AGN unification and evolution \citep{2013ApJ...777...86L}, and thus non-torus obscuration should be taken into account in these studies. 

%

\subsection{Comparison with previous observational and theoretical work}
The existence of large-scale obscuring structures at the centre of AGN has been suggested from statistical analysis of samples of nearby Seyfert and LLAGNs by \cite{1995ApJ...441...96M} and \cite{1995ApJ...454...95M}. The authors find a systematic absence of   type 1 sources in edge-on systems, which suggests the existence of nuclear obscuring material at large scales: a circumnuclear torus at roughly 100 pc scale coplanar with the galaxy disk plane. In this work we present direct observational evidence of large-scale obscuring material but rather randomly distributed about the centre: the nucleus obscuration is fortuitous, produced by a filament or lane crossing the line of sight to the central engine.
Thus, the extended ionised gas should also be randomly distributed and anti-correlated in position with that of the filaments and lanes, as it is indeed found. If these large-scale obscuring structures are common to AGN in general, extended collimated ionised gas is expected to be rare. We would like to remark here the over interpretation of cone-like morphologies. For example, some of the objects studied in this work may look like having a cone-like or collimated ionised gas morphology, e.g. NGC\,1386 or NGC\,3169, yet one can see from the respective figures that gas is also seen through holes in the dust away from the main North-South ionised gas structure, in NGC\,1386, and that the opening angle of the one side ionised gas region is almost 150 degrees, in NGC\,3169.

The morphology of the central obscuring structure here reported ranges from rather chaotic to a well-ordered nuclear spiral or disk-like structure. Nuclear dust spirals have been seen in multiple \textit{HST} studies of nearby active and non-active galaxies using the same technique, optical-IR colour images (see \citealt{2003ApJ...589..774M} and references therein). In this work we are dealing with a smaller subsample but the level of detail achieved is superior. The difference resides on  the use of relatively higher angular scales and of much cleaner PSFs provided by the AO IR images. This allows us to get further into the centre with greater detail than previously done  and, thanks to the accurate astrometry performed, to precisely place the location of the dust relative to the nucleus so that a direct cause of the nucleus obscuration is discovered.  

The existence of a pc-scale nuclear torus is convincingly argued by $10\, \rm{\micron}$ interferometric observations in key type 2 objects as NGC\,1068 and Circinus (\citealt{2009MNRAS.394.1325R} and references therein; \citealt{2014A&A...563A..82T} and references therein), but equally compelling are also the interferometric results for NGC\,424 (type 2) and NGC\,3783 (type 1) where the majority of the pc-scale mid-IR emission is found elongated in the polar direction,  i.e. perpendicular to the expected torus location, which directly challenges the standard interpretation that most of the pc-scale mid-IR emission in AGN originates from the torus (\citealt{2012ApJ...755..149H}; \citealt{2013ApJ...771...87H} and references therein). This polar elongation is also seen in Circinus (\citealt{2014A&A...563A..82T} and references therein). In this work we do not trace the torus scales, hence  neither the high column densities $N_\mathrm{H}$ derived from them, at least an order of magnitude higher than those inferred from the dust filaments and lanes discussed in this work, at most 10$^{22}$ cm$^{-2}$. However, the relative location of dust, ionised gas and the nucleus in the cases presented here is compelling enough to cast doubts on the additional requirement for an AGN torus on a general basis. Specifically to the LLAGN presented in this work, the extinctions derived are sufficient to obscure these faint nuclei and
a torus is definitively not required. This conclusion is in line with recent observational evidence of a receding torus in faint nuclei \citep{2013ApJ...763L...1M}, and with theoretical predictions on the inability of these LLAGN to sustain a torus \citep{2006ApJ...648L.101E,2007MNRAS.380.1172H}. 

Putting altogether, one may still think of the torus as being just part of the same much larger scale -hundred pc- filamentary and lane structure shown here, their optical thickness rising at the very centre if these filaments are channels through which material flows from outer regions to fuel the centre. This goes on the same line of reasoning put already forward by e.g. \citet{1997ApJ...476L..67C} on the basis of \textit{HST} colour maps for NGC\,1068.
Comparing with current modeling, we find a striking resemblance of the extinction maps shown here with those produced in the 3D radiative-feedback hydrodynamic models by \citet{2012ApJ...758...66W}. These models explore the scenario by which radiation pressure from the AGN at different Eddington ratios drives a feedback loop of material outflowing from- and returning back to- the centre. When further coupled with radiative transfer on dust (Schartmann et al. 2014, submitted), the resulting simulations reveal a morphology and location of dust -filaments and -lanes reminiscent of the morphology observed in some of these galaxies, albeit on relatively smaller scale sizes, of several tens of pc, as produced by current modeling. Detailed comparisons are discussed in Schartmann et al. 2014 (submitted).

It can be also possible that the radiation pressure exerted by a torus-collimated nuclear radiation sweeps the medium of dust, naturally leading to conical-shape ionised gas morphologies. Simulations by \cite{2011MNRAS.415..741S} show that the radiation pressure exerted by the powerful nucleus of NGC 1068 is effective in removing dust for gas column densities below a few $10^{23}\, \rm{cm^{-2}}$ (their Fig.\,18), which is well below the column densities inferred from this work (Table\,\ref{extinction}).

\subsection{On the BLR detection in the IR}
One may next wonder whether these filaments are sufficient to hide the broad-line region (BLR), specifically in the IR. It should be noted that BLR searches in the IR are complicated as detections may be hampered by limited angular resolution and a dominating continuum light, either from the host galaxy with low spatial resolution or from the bright nucleus unveiled in the IR with high angular scales. This reduces dramatically the contrast of, specially, broad emission lines and, in particular, of Br$\gamma$, the most common IR BLR tracer, which is furthermore the third in the Bracket series, much fainter than the H$\alpha$ lines.

Broad Br$\gamma$ line emission is undetected in all the galaxies 
 (\citealt{2011ApJ...739...69M}; \citealt{2002A&A...396..439L}; \citealt{2001ApJS..136...61S}; own SINFONI AO data) except for MCG-05-23-016 (FWHM$_{Br\gamma} \sim 2000\, \rm{km\,s^{-1}}$, from our unpublished VLT / SINFONI-AO data, also \citealt{2002A&A...396..439L}), the Seyfert 2 nuclei with the lowest $A_V$ in the  sample. The second galaxy with the lowest
 $A_V$, Sombrero, does not have broad $Br\gamma$ yet it has already been identified in the literature as a genuine  "BLR-free" AGN because of its low $N_\mathrm{H}$, bright UV nucleus and power-law optical continuum (\citealt{2010ApJ...714..115S}; \citealt{2012JPhCS.372a2006F}; Fig.~\ref{NGC4594} this work). 

Variability of the nuclear luminosity might be also a useful test to check whether broad emission lines emerge in periods of high-nuclear activity. However, this emission might also be produced by a supernova event in the crowded star-forming regions located at the nucleus of this galaxy (see Fig.\,\ref{NGC7582}; also \citealt{1999ApJ...519L.123A}).

In summary, the BLR is detected for the Seyfert 2 nucleus less affected by dust absorption, MCG-05-23-016, in agreement with the nuclear dust structure in this object being rather transparent to the nucleus emission in the line of sight. And, aside Sombrero, the BLR remains undetected in the IR in the other studied galaxies although the low $A_\mathrm{V}$ values we measure are insufficient to hide it in the IR. While this could be the indication that a further pc-scale obscuring structure is needed, we find the IR non-detection inconclusive for three reasons: first, the unreliability of Br$\gamma$ as a BLR tracer even at high angular scales; second, nuclear variability (e.g., \citealt{2002A&A...396..439L}), as is the case of the extremely bright Seyfert 2 NGC\,7582 in which  broad H$\alpha$ has been detected on a temporary basis (\citealt{1999ApJ...519L.123A}); third,  the possibility that the non-detection indeed discloses a de-facto absence of the BLR in the LLAGN sources, as suggested on theoretical grounds (\citealt{2009ApJ...701L..91E}) and on modeling of their SEDs (\citealt{2012JPhCS.372a2006F}, 2014 in preparation).

\section{Conclusions and final remarks}\label{conclusions}
This work shows that IR observations at the key wavelength of $2\, \rm{\micron}$ combined with high angular resolution are crucial for unambiguously disclosing and very precisely locating obscured nuclei. We use this fact in a number of prototype type 2 nearby AGN, including Seyferts and LLAGN, to identify and position the nucleus in optical images. The procedure involves accurate astrometry between optical \textit{HST} and IR VLT AO images in which several point-like sources common to the FoV to all images could be identified and their centroid positions determined with a precision of a few tens of mas. The final image registration accuracy is better than $30\, \rm{mas}$. Overall, the optical peak, which is often assumed to be the AGN counterpart in this range, is located a few tens of pc away from the IR-nucleus and is thus unrelated to the true galaxy nucleus. The nature of the optical peak is varied: it often relates to a bright blob in the extended ionised gas, a young stellar cluster in cases where nuclear star formation is present, or just the maximum of the stellar brightness profile. This misidentification sets a warning on more general statistical classification of AGN, in particular of type 2 and transient objects on the basis of optical spectra alone (e.g. BPT diagrams) and on the misleading conclusions that may be derived from these (e.g. ionisation level, bolometric luminosities, or SEDs).

To asses the nature of the nucleus obscuration, dust extinction maps (optical--IR colour maps) of the central kpc region of each galaxy are constructed using the registered image set. Because of the wide wavelength range and equivalent angular scales between the optical and IR images used, the maps show with an unprecedented detail the morphology and location of dust from the kpc to the central few pc in these galaxies. We find that in all cases the nucleus is strategically covered by a large scale --a few hundred pc-- dust filament or diffuse dust lane. The extinction $A_V$ derived from these structures, as inferred from the optical--IR colour maps, is moderate ($A_V \sim 3$--$6\, \rm{mag}$). However, the sharp disappearance in optical emission lines, specifically H$\alpha$, in regions next to the dust location suggests that these extinctions may be underestimated. Only in two cases is $A_V$ $\sim$ $1\, \rm{mag}$, in agreement with the fully visible nucleus up to the UV shown in one case (Sombrero) and partially visible in the second (MCG-05-23-016). These central dust filaments are definitely a source of nuclear attenuation and in some cases it may be the major, if not the only, cause of the nucleus obscuration. Albeit the apparent moderate extinctions, the presence of dust on large scales in the centre of all these galaxies, its morphology and strategic location just in front of the nucleus questions whether ultimately the putative AGN torus is responsible for the obscuration of the nuclear emission.
 
Using the same image registration accuracy and angular scale, the location of the ionised gas relative to that of the nucleus and dust is determined (specifically H$\alpha$ and/or the coronal [\textsc{Si\,vii}] $2.48\, \rm{\micron}$ line, the latter being an unequivocal AGN tracer). H$\alpha$ is found sharply anti-correlated in position with that of the observed dust filaments and lanes. The preferential direction in which this gas extends in some of these galaxies, sometimes defining a conical morphology, appears not to be caused by a nuclear collimation --torus, as usually thought-- but by the location of the dust on larger scales. This is however not the case for the two cases (Sombrero and MCG-05-23-016) for which $A_V$ $\sim 1\, \rm{mag}$. Intriguingly, in Sombrero H$\alpha$ and dust show the same overlapping morphology, much suggesting an edge-on star-forming disk of which no evidence has so far been reported. Conversely, the [\textsc{Si\,vii}] $2.48\, \rm{\micron}$ coronal line, less subjected to dust extinction, shows a rather homogeneous morphology --less patchy than is often the case for the optical lines-- and a distribution isotropic in some cases or extending, coincidentally, along the galaxy mayor axis in others, suggesting an emission disk in the plane of the galaxy.

10 $\mu$m interferometric observations are revealing the existence of a pc-scale torus in some objects whereas challenging its existence in others. The precise determination of dust filaments and lanes passing in front of the nucleus and their strict anti-correlation with the ionised gas found in most of these galaxies challenges the requirement for a torus as key structure to hide the nucleus and to collimate the nuclear radiation. 
Specifically to the LLAGN class, a torus is not needed. Still, one may interpret these central dust structures as the feeding channels through which material flows from outer scales - few 100 pc -  in the galaxy to fuel the centre.
An alternative interpretation for the origin of these filaments arises from the results on current 3D radiative-feedback hydrodynamic simulations coupled with dust radiative transfer, which show a remarkable similarity in the morphology and location of dust as those in the extinction maps shown here. In this case, radiation pressure from the AGN drives a feedback loop of material outflowing from- and returning back to- the centre. 

We call the attention on the striking morphology of the dust-filament complex crossing the nucleus of NGC\,1386, reminiscent of the infinite-symbol shape. A very similar shape is shown by the central filamentary structure enclosing the supermassive black hole Sgr A* in the Milky Way, also extending over similar scales of $\sim$100 pc. The presence of this structure in the Milky Way is raising enormous attention in the community, its origin and role remains unknown. It is however intriguing that an equivalent structure is seen at the centre of other LLAGN, which opens the possibility of being the way in which low-accreting black holes are fueled. 

Finally, this work provides the most accurate to our knowledge nucleus positioning of NGC\,1068 relative to optical images, located at the South vertex of the so-called cloud B with a precision of about $30\, \rm{mas}$. This position is displaced from all previous determinations of the nucleus location, yet it is consistent, within the literature reported errors, with the position of the symmetry centre of the UV polarisation pattern.

\section*{Acknowledgments}
This research is supported by Spanish grant under contract PN-AYA2011-25527.
M.\,A. Prieto acknowledges the hospitality of the Max-Planck-Institut f\"ur Extraterrestrische Physik where a large fraction of this work was done, and especially the CAST group for stimulant discussions. J.\,A. acknowledges the hospitality of Instituto de Astrof\'isica de Canarias. M.S. was supported by the Deutsche Forschungsgemeinschaft priority program 1573 (``Physics of the Interstellar Medium'').
\bibliographystyle{mn2e} 
\bibliography{references_may2014}


\begin{figure*}
 \includegraphics[width=0.85\textwidth]{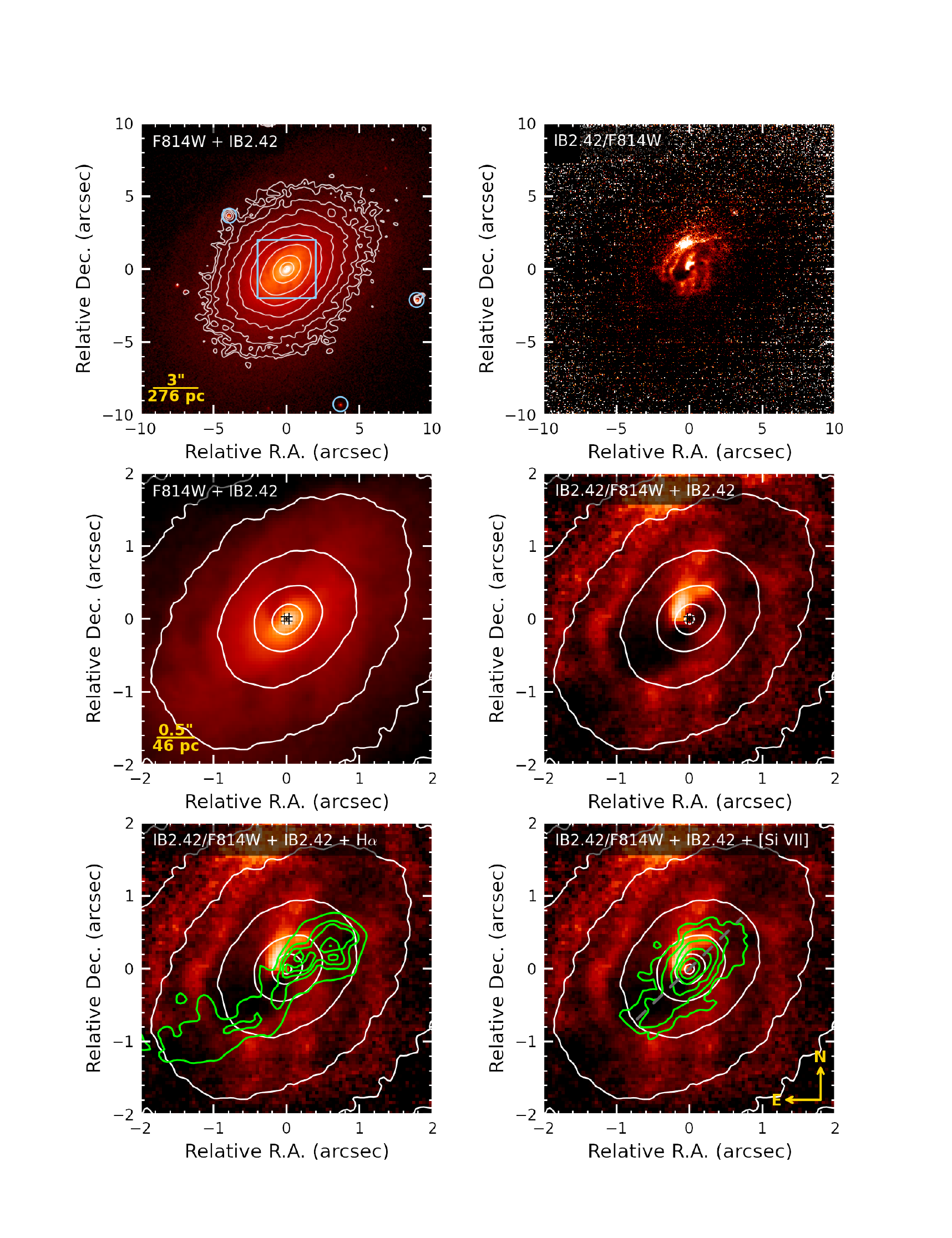}
 \protect\caption[ESO428]{ESO\,428-G14. \textbf{Top left}: \textit{HST}/F814W image with \textit{K}-band continuum contours in white (from IB2.42 narrow-band filter). The FoV is $20\arcsec \times 20\arcsec$. Blue circles mark the position of the point-like sources used for image alignment. The inner $4\arcsec \times 4\arcsec$ region (blue square) is shown in detail in the middle and bottom panels. \textbf{Top right:} IB2.42/F814W ratio or dust map with the same FoV as the previous panel. \textbf{Middle left:} F814W image with IB2.42 contours in white. \textbf{Middle right:} IB2.42/F814W ratio with IB2.42 contours in white. The position of the nucleus and its error is marked with a cross in the middle panels. \textbf{Bottom left:} IB2.42/F814W ratio with IB2.42 contours in white and H$\alpha$ emission line contours in green. \textbf{Bottom right:} IB2.42/F814W ratio with IB2.42 contours in white and [\textsc{Si\,vii}] coronal emission line contours in green. The orientation of the radio jet \citep[e.g.][]{1999ApJ...516...97N} is shown with a dashed grey line. North is up and East is to the left.}
 \label{ESO428}
\end{figure*}

 \begin{figure*}
 \includegraphics[width=0.85\textwidth]{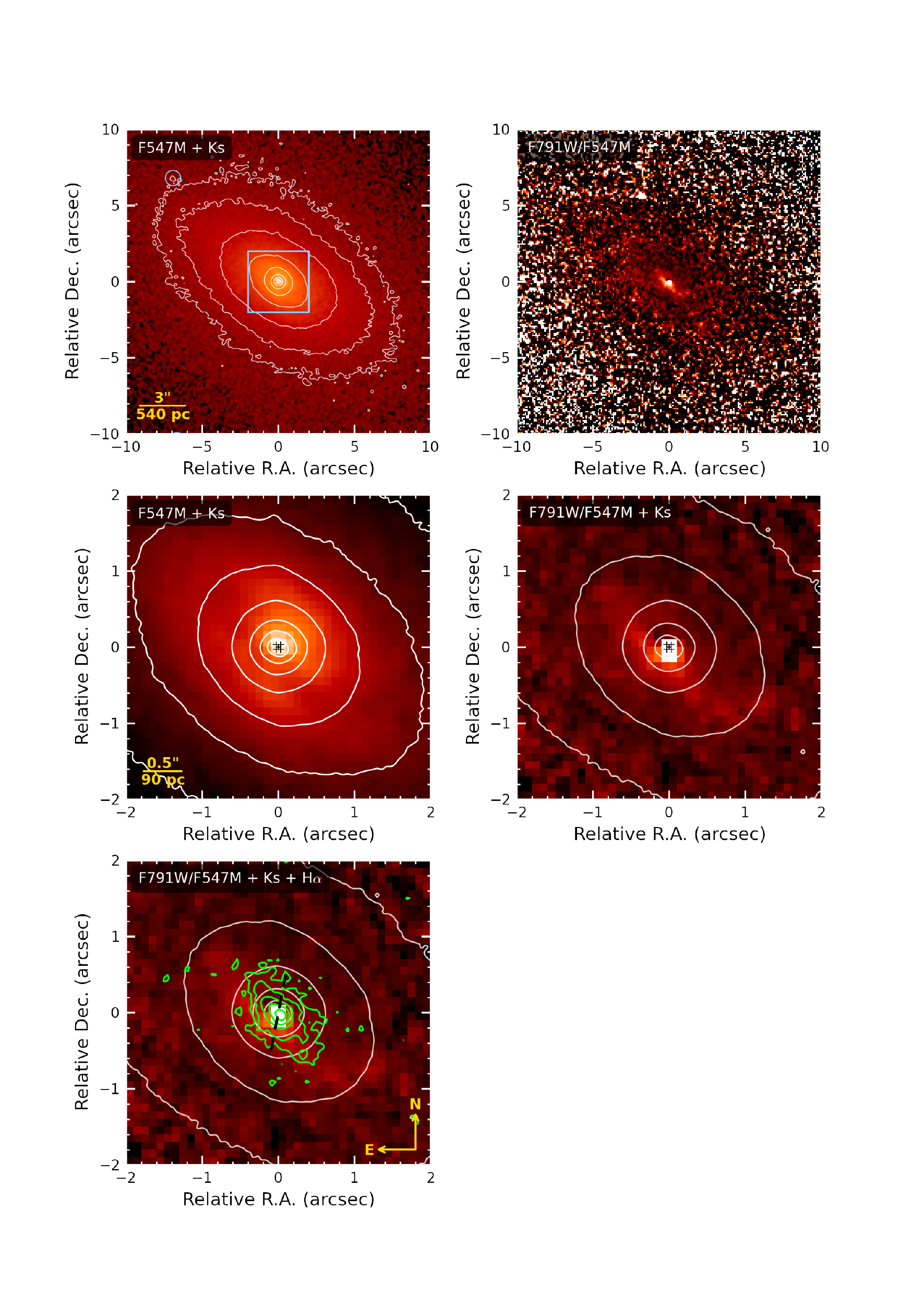} 
 \protect\caption[MCG]{MCG-05-23-016. \textbf{Top left:} \textit{HST}/F547M image with \textit{Ks}-band continuum contours in white. The FoV is $20\arcsec \times 20\arcsec$. Blue circles mark the position of the point-like sources used for image alignment. The inner $4\arcsec \times 4\arcsec$ region (blue square) is shown in detail in the middle and bottom panels. \textbf{Top right:} F791W/F547M ratio or dust map with the same FoV as the previous panel. \textbf{Middle left}: F547M image with \textit{Ks}-band contours in white. \textbf{Middle right:} F791W/F547M ratio with \textit{Ks} contours in white. The position of the nucleus and its error is marked with a cross in the middle panels. \textbf{Bottom left:} F791W/F547M ratio with \textit{Ks} contours in white and H$\alpha$ contours in green. The orientation of the radio jet \citep[e.g.][]{2009ApJ...703..802M,2010MNRAS.401.2599O} is indicated with a dashed black line. North is up and East is to the left.}
 \label{MCG}
\end{figure*}

 \begin{figure*}
 \includegraphics[width=0.85\textwidth]{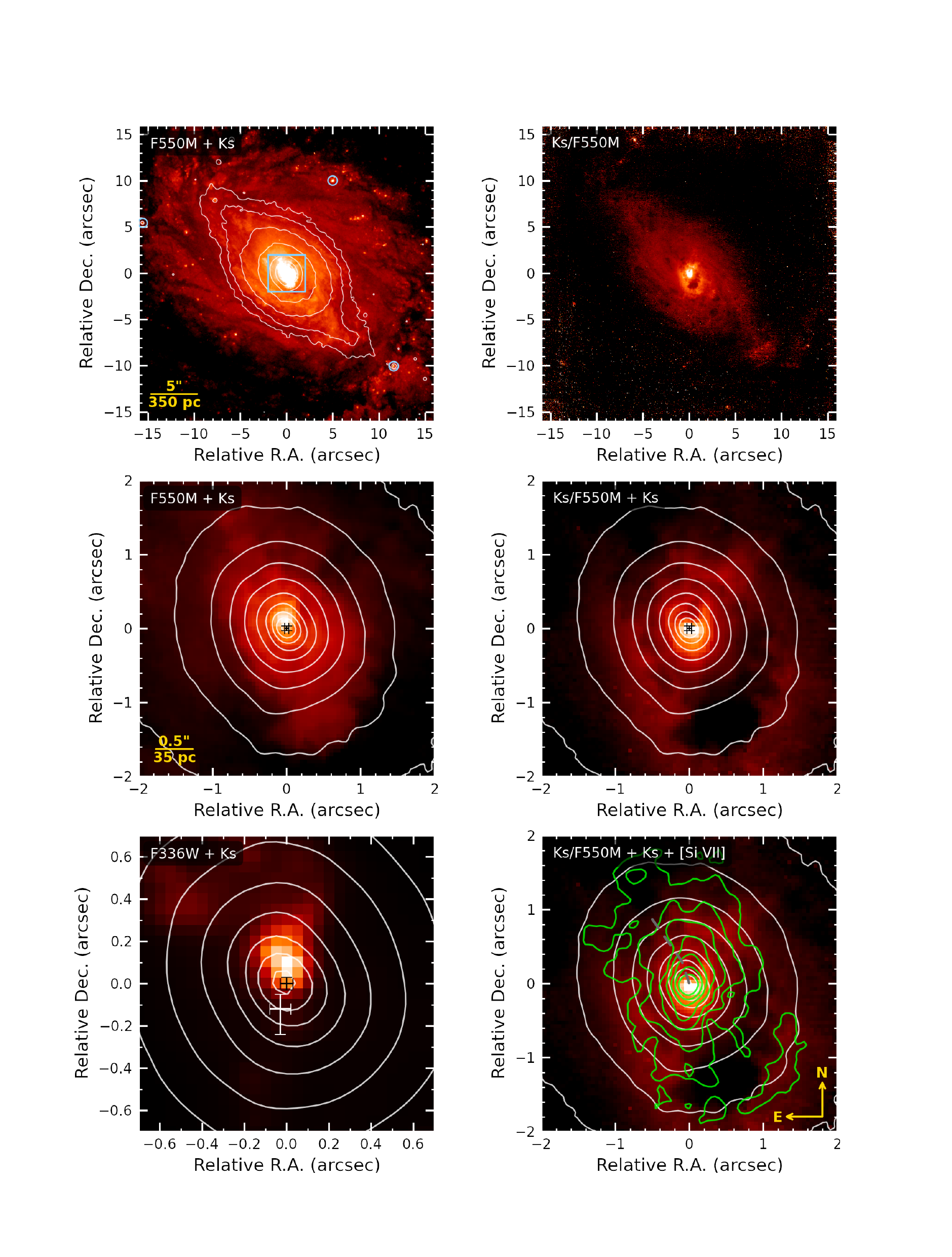}
 \protect\caption[NGC1068]{NGC\,1068. \textbf{Top left:} \textit{HST}/F550M image with \textit{Ks}-band continuum contours in white. The FoV is $32\arcsec \times 32\arcsec$. Blue circles mark the position of the point-like sources used for image alignment. The inner $4\arcsec \times 4\arcsec$ region (blue square) is shown in detail in the middle and bottom right panels. \textbf{Top right:} \textit{Ks}/F550M ratio or dust map with the same FoV as the previous panel. \textbf{Middle left:} F550M image with \textit{Ks} contours in white. \textbf{Middle right:} \textit{Ks}/F550M ratio with \textit{Ks} contours in white. The position of the nucleus and its error is marked with a cross in the middle panels. \textbf{Bottom left:} \textit{HST}/F336W image showing the UV cloud B \citep{1991ApJ...369L..27E} with \textit{Ks}-band continuum contours in white. The white cross marks the position of the nucleus from polarimetry measurements \citep{1999ApJ...518..676K}. Our position of the nucleus and its error is marked with a black cross. \textbf{Bottom right:} \textit{Ks}/F550M ratio with \textit{Ks} contours in white and [\textsc{Si\,vii}] coronal emission line contours in green. The orientation of the radio jet \citep{1996ApJ...464..198G} is indicated with a dashed grey line. North is up and East is to the left.}
 \label{NGC1068}
\end{figure*}

 \begin{figure*}
 \includegraphics[width=0.85\textwidth]{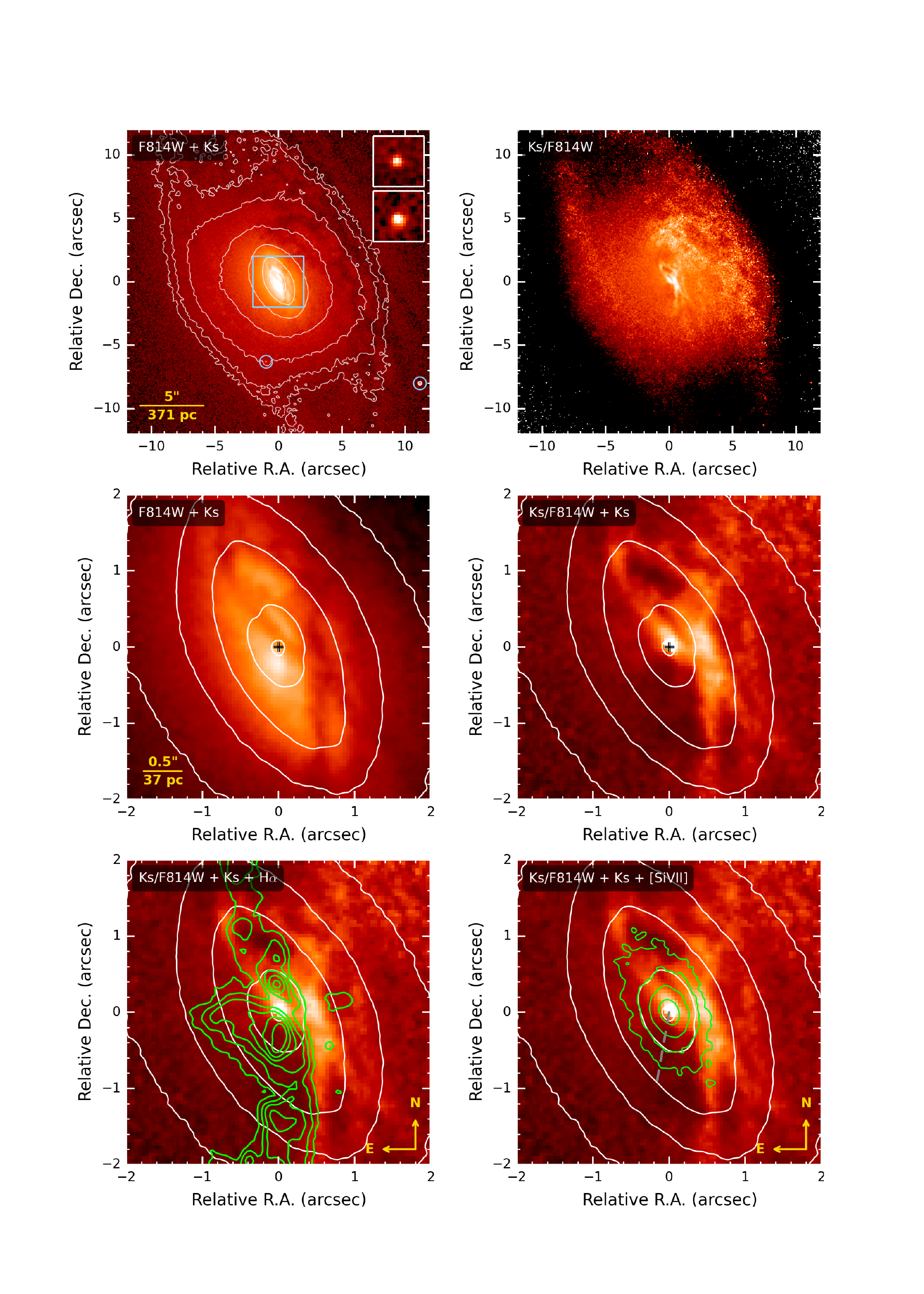}
 \protect\caption[NGC1386]{NGC\,1386. \textbf{Top left:} \textit{HST}/F814W image with \textit{Ks}-band continuum contours in white (from IB2.42 narrow-band filter). The FoV is $24\arcsec \times 24\arcsec$. Blue circles mark the position of the point-like sources used for image alignment, which are also shown in the two inner panels at the top-right corner. The inner $4\arcsec \times 4\arcsec$ region (blue square) is shown in detail in the middle and bottom panels. \textbf{Top right:} \textit{Ks}/F814W ratio or dust map with the same FoV as the previous panel. \textbf{Middle left:} F814W image with \textit{Ks} contours in white. \textbf{Middle right:} \textit{Ks}/F814W ratio with \textit{Ks} contours in white. The position of the nucleus and its error is marked with a cross in the middle panels. \textbf{Bottom left:} \textit{Ks}/F814W ratio with \textit{Ks} contours in white and H$\alpha$ contours in green. \textbf{Bottom right:} \textit{Ks}/F814W ratio with \textit{Ks} contours in white and [\textsc{Si\,vii}] coronal emission line contours in green. The orientation of the radio jet \citep{1999ApJ...516...97N} is plotted with a dashed grey line. North is up and East is to the left.}
 \label{NGC1386}
\end{figure*}

\begin{figure*}
 \includegraphics[width=0.85\textwidth]{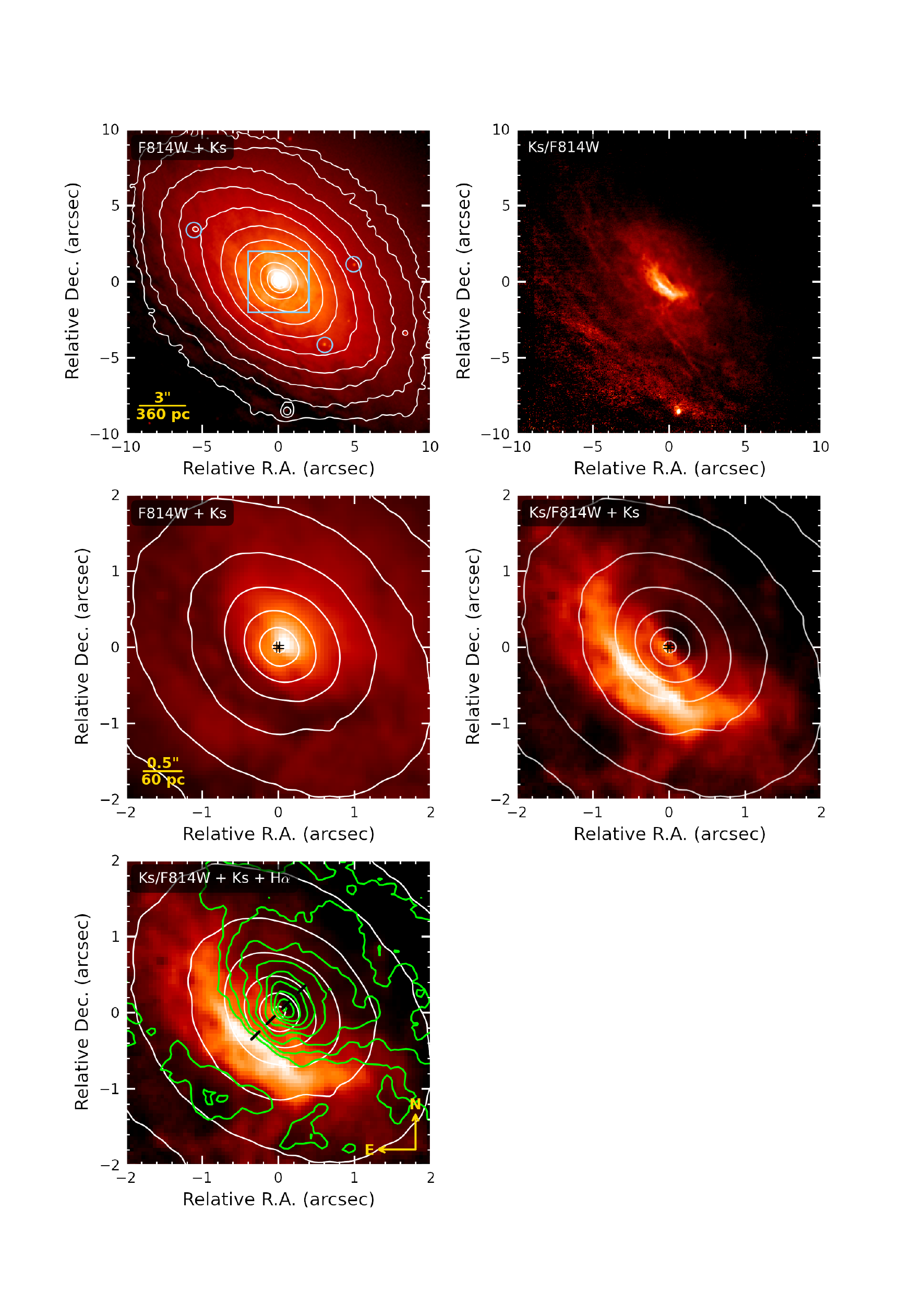}
 \protect\caption[NGC3169]{NGC\,3169. \textbf{Top left:} \textit{HST}/F814W image with \textit{Ks}-band continuum contours in white. The FoV is $20\arcsec \times 20\arcsec$. Blue circles mark the position of the point-like sources used for image alignment. The inner $4\arcsec \times 4\arcsec$ region (blue square) is shown in detail in the middle and bottom panels. \textbf{Top right:} \textit{Ks}/F814W ratio or dust map with the same FoV as the previous panel. \textbf{Middle left:} F814W image with \textit{Ks} contours in white. \textbf{Middle right:} \textit{Ks}/F814W ratio with \textit{Ks} contours in white. The position of the nucleus and its error is marked with a cross in the middle panels. \textbf{Bottom left:} \textit{Ks}/F814W ratio with \textit{Ks} contours in white and H$\alpha$ contours in green. The orientation of the radio jet \citep{2006A&A...451...71F} is shown with a dashed black line. North is up and East is to the left.}
 \label{NGC3169}
\end{figure*}

 \begin{figure*}
 \includegraphics[width=0.85\textwidth]{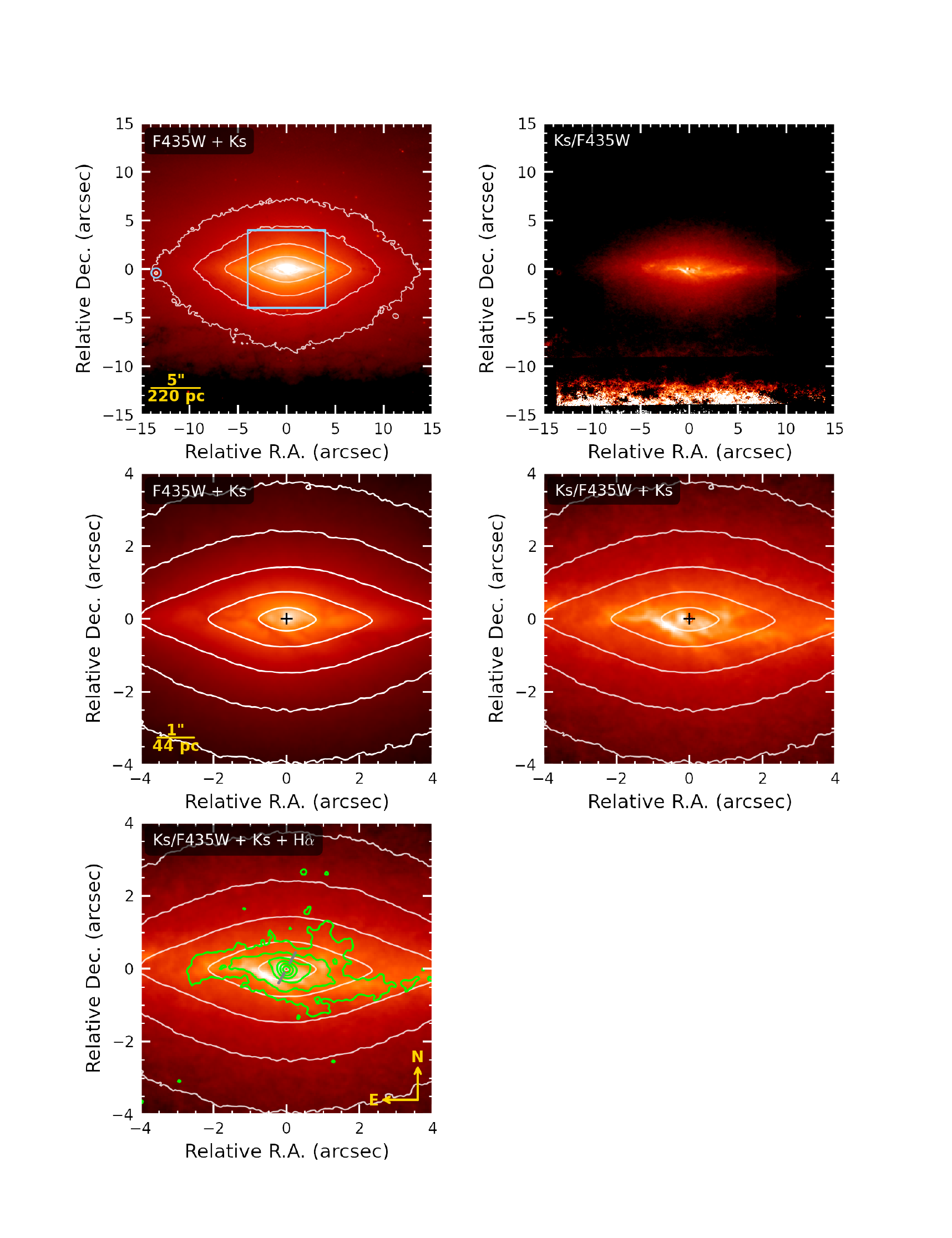}
 \protect\caption[NGC4594]{Sombrero Galaxy (NGC\,4594). \textbf{Top left:} \textit{HST}/F435W image with \textit{Ks}-band continuum contours in white. The FoV is $30\arcsec \times 30\arcsec$. Blue circles mark the position of the point-like sources used for image alignment. The inner $8\arcsec \times 8\arcsec$ region (blue square) is shown in detail in the middle and bottom panels. \textbf{Top right:} \textit{Ks}/F435W ratio or dust map with the same FoV as the previous panel. \textbf{Middle left}: F435W image with \textit{Ks} contours in white. \textbf{Middle right:} \textit{Ks}/F435W ratio with \textit{Ks} contours in white. The position of the nucleus and its error is marked with a cross in the middle panels. \textbf{Bottom left:} \textit{Ks}/F435W ratio with \textit{Ks} contours in white and H$\alpha$ contours in green. The orientation of the radio jet \citep[]{2006AJ....132..546G,2014ApJ...787...62M} is indicated with a dashed grey line. North is up and East is to the left.}
 \label{NGC4594}
\end{figure*}

 \begin{figure*}
 \includegraphics[width=0.85\textwidth]{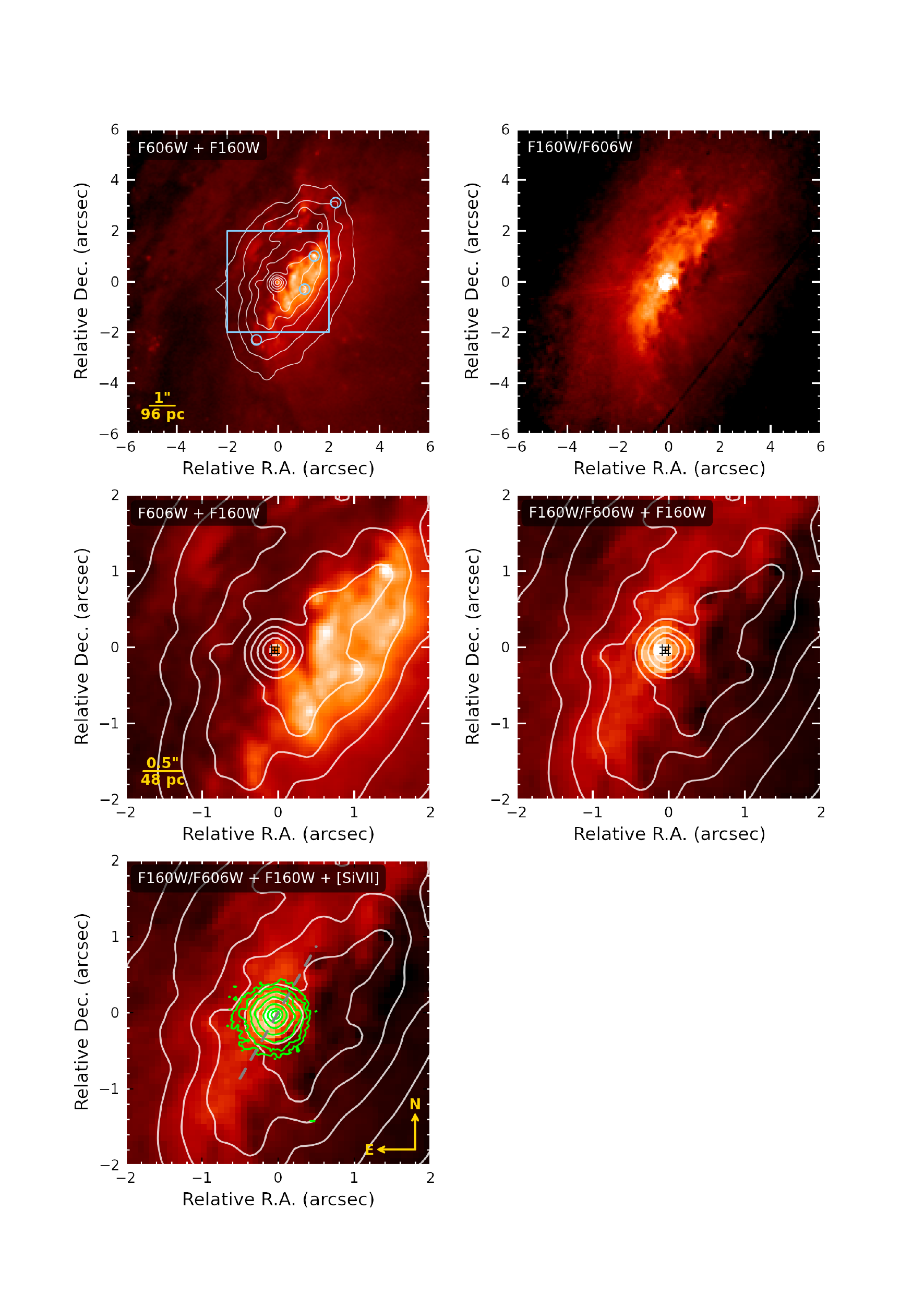}
 \protect\caption[NGC7582]{NGC\,7582. \textbf{Top left:} \textit{HST}/F606W image with NICMOS/F160W continuum contours in white. The FoV is $12\arcsec \times 12\arcsec$. Blue circles mark the position of the point-like sources used for image alignment. The inner $4\arcsec \times 4\arcsec$ region (blue square) is shown in detail in the middle and bottom panels. \textbf{Top right:} F160W/F606W or dust map with the same FoV as the previous panel. \textbf{Middle left:} F606W image with F160W contours in white. \textbf{Middle right:} F160W/F606W ratio with F160W contours in white. The position of the nucleus and its error is marked with a cross in the middle panels. \textbf{Bottom left:} F160W/F606W ratio with F160W contours in white and [\textsc{Si\,vii}] coronal emission line contours in green. The orientation of the narrow-line region \citep{1984ApJ...285..439U,1999A&AS..137..457M} is shown with a dashed grey line. North is up and East is to the left.}
 \label{NGC7582}
\end{figure*}

 \begin{figure*}
 \includegraphics[width=0.85\textwidth]{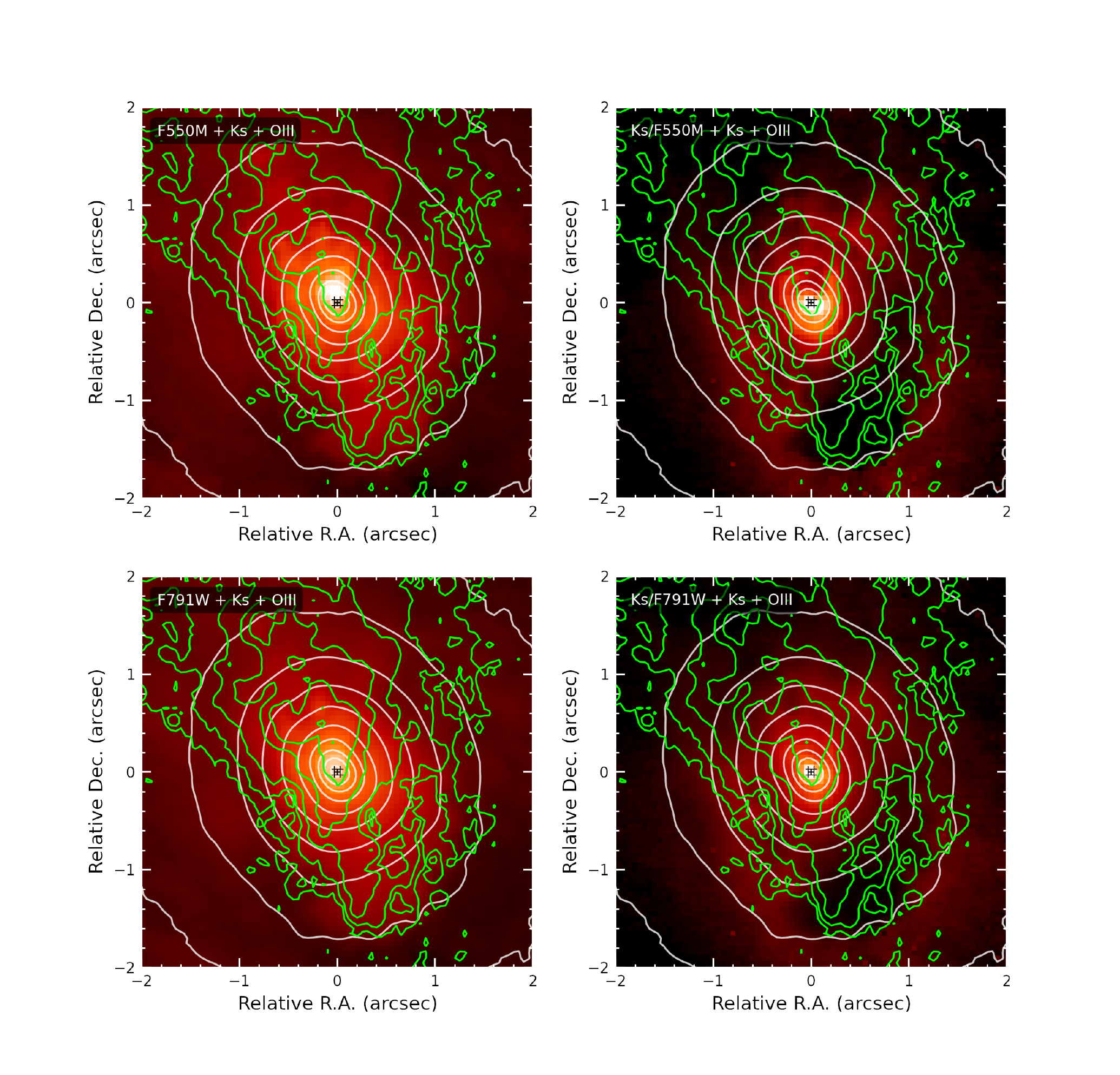}
 \protect\caption[NGC1068_OIII]{NGC\,1068. \textbf{Top left, right:} \textit{HST}/F550M image and \textit{Ks}/F550M ratio or dust map, respectively, with a FoV of $4\arcsec \times 4\arcsec$. \textbf{Bottom left, right:} \textit{HST}/F791W image and \textit{Ks}/F791W ratio or dust map, respectively, with a FoV of $4\arcsec \times 4\arcsec$. In all panels the \textit{Ks}-band continuum contours are shown in white and the [\textsc{O\,iii}] emission line contours in green. The lowest [\textsc{O\,iii}] contour corresponds to $3\sigma$. The position of the nucleus and its error is marked with a cross in all panels. North is up and East is to the left.}
 \label{NGC1068_OIII}
\end{figure*}

\begin{figure*}
\includegraphics[width=\textwidth]{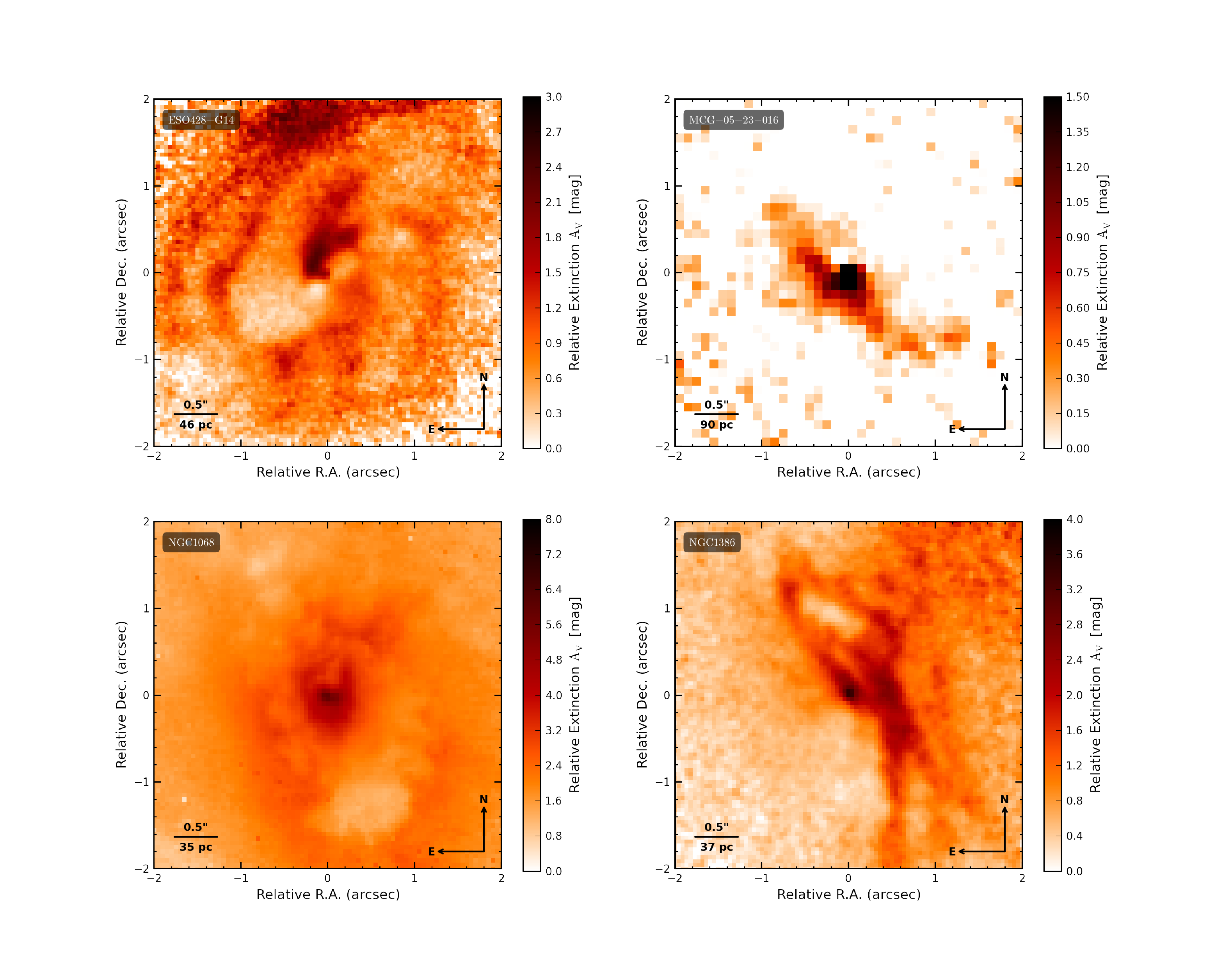}
\protect\caption[extinction1]{Extinction maps of ESO\,428-G14, MCG-05-23-016, NGC\,1068 and NGC\,1386 derived using the same dust maps for each object as in the corresponding individual Figs\,\ref{ESO428}--\ref{NGC1386}. The reference level is taken from a dust-free region away from the filamentary structures and within a kpc radius from the centre. North is up and East is to the left.}
 \label{extinction1}
\end{figure*}

\clearpage

\begin{figure*}
\includegraphics[width=\textwidth]{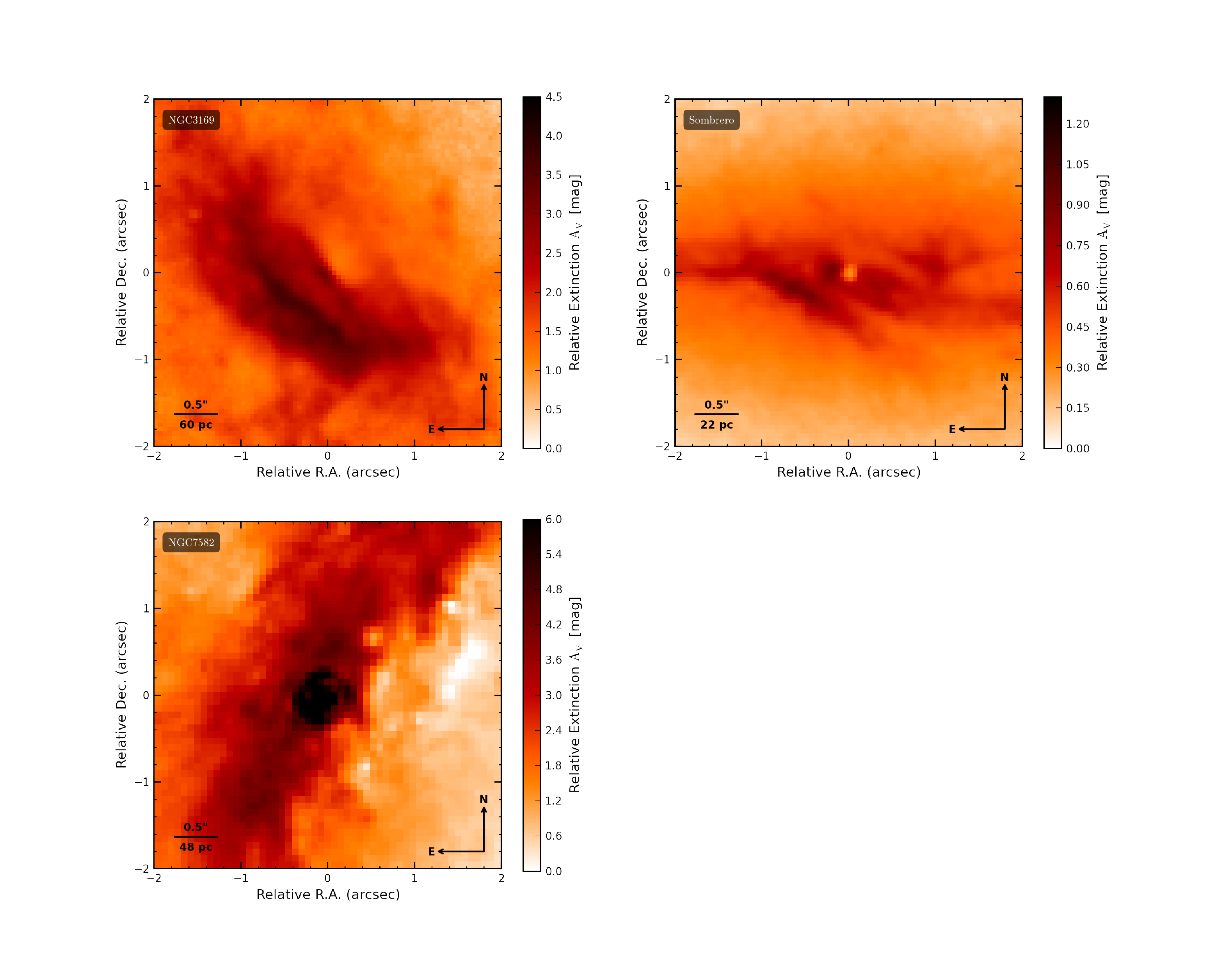}
\protect\caption[extinction1]{Extinction maps of NGC\,3169, Sombrero and NGC\,7582 derived using the same dust maps for each object as in the corresponding individual Figs\,\ref{NGC3169}--\ref{NGC7582}. The reference level is taken from a dust-free region away from the filamentary structures and within a kpc radius from the centre. North is up and East is to the left.}
 \label{extinction2}
\end{figure*}

\appendix
\cleardoublepage

\section{Discussion on individual objects} \label{individual}
\subsection*{ESO\,428-G14}
The registration was based on centroid measurements of three sources common to the FoV to all images (Fig.\,\ref{ESO428}, top left panel). The nucleus is obscured in the \textit{HST}/F814W filter. The optical peak is offset by $50 \pm 27\, \rm{mas}$ from the IR peak, this being measured in the line-free narrow-band filter VLT-NaCo/IB$2.42\, \rm{\micron}$ and identified as the nucleus. This identification is further confirmed by the location of the peak of the [\textsc{Si\,vii}] $2.48\, \rm{\micron}$ coronal emission (Fig.\,\ref{ESO428}, bottom right panel), which falls on top of the IR peak.
The colour map IB2.42/F814W (Fig.\,\ref{ESO428}) indicates that the nucleus obscuration is presumably caused by the intersection of two narrow filaments which are spiraling towards the centre. The map further shows other dust filaments and lanes within the central few hundred pc in the galaxy, spiraling towards the centre. On the kpc scales, both filaments and lanes are traced up to the edges of the visible body of the galaxy in the image (Fig.\,\ref{ESO428}, top right), at $\sim 1.5\, \rm{kpc}$ radius. The largest extinction of these filaments, $A_V \sim 3\, \rm{mag}$, is found at the section crossing the nucleus, and it is presumably its major source of obscuration.
The H$\alpha$ emission is strategically located within the limits of the dust filaments and fills-up two major dust-free cavities left by them at both sides of the nucleus (Fig.\,\ref{ESO428}, bottom left panel). The coronal [\textsc{Si\,vii}] $2.48\, \rm{\micron}$ extends more uniformly also at both sides of the nucleus and along the same direction as that of H$\alpha$ and radio. This direction is that of the major axis of the galaxy (Fig.\,\ref{ESO428}, bottom right panel).

\subsection*{MCG-05-23-016}
The nucleus is obscured below $5500\, \rm{\AA}$ but becomes visible from $7900\, \rm{\AA}$ onward. It reaches its maximum at $2\, \rm{\micron}$ and keeps as a prominent source up to $20\, \rm{\micron}$, the longest wavelength at which high-angular-resolution images are available \citep{2010MNRAS.402..879R}. An offset of $80 \pm 30\, \rm{mas}$ is determined between the optical peak --in \textit{HST}/F547M filter-- and the IR peak in the NaCo/\textit{Ks}-band (Fig.\,\ref{MCG}, middle left panel). The dust extinction map (F791W/F547M, Fig.\,\ref{MCG}, middle right panel) shows a contrasted point-like source at the centre that coincides in position with the photometric peak in \textit{HST}/F791W and NaCo \textit{Js}- (not shown) and \textit{Ks}-band images within our astrometric errors (Table\,\ref{filtros}). This source, assumed to be the nucleus, seems partially obscured by either an edge-on ring or a filament: in the extinction maps in Fig.\,\ref{MCG} one can see the nucleus slightly shifted up North from this structure thus causing a partial obscuration. This structure has a length of $\sim 2\arcsec$ ($360\, \rm{pc}$), it is placed along the major axis of the galaxy and it is seen in all dust extinction maps constructed (\textit{K}-band/F547M and NaCo \textit{Ks}/\textit{Js}, the later not shown). There is no detection of a larger scale dust structure connected to this $2\arcsec$ region (Fig.\,\ref{MCG}, top right panel), contrary to what is seen in other galaxies in this study. The inferred extinction is very low, $A_V$ $\sim 1\, \rm{mag}$, consistent with the partial obscuration of the source. 
 The \textit{HST}/H$\alpha$ peak emission falls on top of the NaCo/IR and \textit{HST}/F791W peaks (Fig.\,\ref{MCG}, bottom). The H$\alpha$ emission --whose morphology is similar to that in [\textsc{O\,iii}] \citep{2000ApJS..128..139F}-- extends to both sides of the nucleus and does not show any evidence of light suppression by the dust feature. Instead, both the H$\alpha$ and [\textsc{O\,iii}] emission seem to be placed along the major axis of the galaxy, perhaps in a disk similar to what is seen in Sombrero (Fig.\,\ref{NGC4594}, bottom), which is consistent with the inferred low $A_V$.

\subsection*{NGC\,1068}
The nucleus is well known to be obscured at all optical wavelengths (e.g. \citealt{1997ApJ...476L..67C}) including at $1\, \rm{\micron}$ (\citealt{2010MNRAS.402..724P}). It becomes visible at the \textit{H}-band and reaches the brightness peak from \textit{K}-band onward up to at least $20\, \rm{\micron}$, the limit of available high-angular-resolution images (\citealt{2004A&A...417L...1R}; \citealt{2010MNRAS.402..724P}). 
The image registration is based on the centroid position of three point-like sources identified in the common FoV of the NaCo/\textit{Ks}-band image and optical \textit{HST} images. These are located between $5\arcsec$ to $10\arcsec$ from the nucleus (Fig.\,\ref{NGC1068}, top left panel). The astrometry procedure allows us to obtain the most accurate determination to date of the nucleus location of NGC\,1068 in the optical with an uncertainty of $\sim 30\, \rm{mas}$ (Table\,\ref{filtros}). Specifically, the astrometry was done between the \textit{HST}/F550M and F791W images and the NaCo/\textit{Ks}-band image. The photometric peak in NaCo/\textit{Ks} band is found to be shifted with respect to the optical peak by $110 \pm 30\, \rm{mas}$ in the \textit{HST}/F550M filter and by half that value, $50 \pm 30\, \rm{mas}$, in the \textit{HST}/F791W filter. The different shift is a consequence of lower extinction towards longer wavelengths and the complex morphology of the nuclear region of NGC\,1068.

The accuracy of this astrometry is better than previous estimates based on optical-IR images. The offset found is a factor three to four smaller than previously reported using also \textit{HST} optical and AO \textit{Ks}-band images: $0\farcs28 \pm 0\farcs05$-South, $0\farcs08 \pm 0\farcs05$-West (\citealt{1997A&A...320..399M}, using \textit{HST} I-band); $0\farcs41 \pm 0\farcs10$-South and $0\farcs23 \pm 0\farcs10$-West (\citealt{1997ApJ...490..238T}, using \textit{V}-band). The reason for the discrepancy is the method used, which relies on the precision of telescope coordinates of stars taken in separate frames from that of the galaxy. On a different approach, based on \textit{HST} imaging polarimetry, \citealt{1997ApJ...476L..67C} determined the location of the source of the scattered radiation assuming that to be the active nucleus. These authors find the nucleus to be $300\, \rm{mas}$ south and $10\, \rm{mas}$ West of the optical continuum peak with an accuracy of $100\, \rm{mas}$. Even within the errors, this position is still further South than that derived from our astrometry by at least $90 \pm 30\, \rm{mas}$. \citet{1999ApJ...518..676K} reviewed the \textit{HST} polarimetry-image analysis and refined the position of the nucleus, to be now located 120$^{+70}_{-120}\, \rm{mas}$ South of the UV peak. This UV peak is located in the so-called ``cloud B" after \citet{1991ApJ...369L..27E}. Its location relative to the NaCo $2\, \rm{\micron}$ peak is shown in Fig.\,\ref{NGC1068} (bottom left) by means of a \textit{HST}/F336W image. The registration of both images using common point-like sources in the FoV yields a shift of $130 \pm 91\, \rm{mas}$ between the UV peak and the $2\, \rm{\micron}$ peak (Table\,\ref{filtros}). Kisimoto et al. centre of polarization, also shown in the figure, is within its relatively large errors in agreement with the $2\, \rm{\micron}$ peak.

In our work, we find that the optical (\textit{HST}/F550M) peak and the \textit{HST}/UV peak coincide within the errors, the optical peak being better defined as it appears more point-like. The central bright $2\, \rm{\micron}$ source in the NaCo images is thus placed at $110 \pm 30\, \rm{mas}$ ($7.7\, \rm{pc}$) of cloud B peak, just at its South vertex (Fig.\,\ref{NGC1068}, bottom left). The alignment of the NaCo \textit{Ks}-band and the [\textsc{Si\,vii}] $2.48\, \rm{\micron}$ coronal emission places the brightest and most collimated [\textsc{Si\,vii}] region on top the \textit{Ks}-band photometric peak (Fig.\,\ref{NGC1068}, bottom right).

The colour map \textit{Ks}-band/F550M (Fig.\,\ref{NGC1068}) shows the more complex morphology among all the galaxies in this study. The nucleus is a highly contrasted source at the centre, coinciding with the \textit{K}-band peak. It is surrounded by a prominent ring with a horn-shape morphology pointing towards the North. This structure extends over $50\, \rm{pc}$ radius from the nucleus and its extinction is $A_V \sim 6\, \rm{mag}$, as measured in the \textit{K}-band/F791W colour map. The coronal [\textsc{Si\,vii}] emission slips in between the two branches of the horn towards the North, as it is also seen in the UV continuum light --cloud B (Fig.\,\ref{NGC1068}, bottom panels). Cloud B has been interpreted as off-scattered light from the nucleus. The location of the nucleus and observed distribution of the central dust, embracing the UV emission, very much illustrate that to be the case. Beyond this horn-like structure, there is a clear light depression in the colour maps: colours become less bluer and the morphology patchy. This is illustrated by a number of darker regions --cavities-- seen in the colour maps, in particular in the large-scale colour map (Fig.\,\ref{NGC1068}, top right panel). One of the largest cavities is clearly seen South of the nucleus, at about $1\arcsec$ radius. 
Further beyond $\sim 1\arcsec$ radius, a second ring-like structure --kind of collar morphology-- with redder colours develops. The estimated extinction in the \textit{K}-band/F791W colour image is about $3\, \rm{mag}$. Two issues should be noted assuming that these various regions just trace the distribution of dust with different optical thickness: (1) although the estimated extinction next to the centre is high, $A_V \sim 6\, \rm{mag}$, the optical ionisation lines [\textsc{O\,iii}] $5007\, \rm{\AA}$ and H$\alpha$ spread everywhere around the nucleus (e.g. Fig.\,\ref{NGC1068}; \citealt{1997ApJ...476L..67C}). However, the outer boundary of their emission seems determined by this collar-like dust structure, as illustrated in Fig.\,\ref{NGC1068_OIII} for the [\textsc{O\,iii}] emission. The [\textsc{O\,iii}] emission spreads all over the central region including the cavity regions but gets suppressed when reaching the collar-like region. (2) H$\alpha$ shows a very similar morphology to [\textsc{O\,iii}], and the IR coronal lines [\textsc{Si\,vi}] $1.96\, \rm{\micron}$ and \textsc{Si\,vii}] $2.48\, \rm{\micron}$ also drop off at the same boundary limits (see Fig.\,\ref{NGC1068}, bottom right panel, and Fig.\,4 in \citealt{2013MNRAS.430.2411M}). 
This anti-correlation in position between dust and ionised line emission argues for the overall ionisation cone shape, well known in NGC\,1068, to be defined by the dust distribution rather than the torus. 
Nevertheless, it is not clear why the optical line emission is not attenuated at the central region where the estimated $A_V$ are large. Projection effects may play a role, or the colour maps do not trace dust but e.g. a change in stellar population within the $0.5\, \rm{kpc}$ radius where we are making the colour comparison. Yet, the [\textsc{O\,iii}] peak emission is definitively shifted from the IR nucleus, by $50 \pm 30\, \rm{mas}$ ($4 \pm\, 2 \rm{pc}$), thus the emission at the very centre gets indeed suppressed, which argues for a pc-scale obscuring structure at the centre.

\subsection*{NGC\,1386}
The nucleus of NGC\,1386 is obscured in all the optical filters (\textit{HST}/F547M, F606W and F814W) and in NaCo/\textit{J}-band images. It appears as an outstanding bright source in NaCo/\textit{K}-band and longward up to $20\, \rm{\micron}$ \citep{2010MNRAS.402..879R}. The separation between the optical and IR peak is $230 \pm 10\, \rm{mas}$ ($17\, \rm{pc}$; see Table\,\ref{filtros}), the largest --in arcsecs-- in the sample. The obscuration seems to be caused by a thin $\sim 0\farcs2$-wide ($15\, \rm{pc}$) dust filament crossing the centre straight away (see \textit{Ks}-band/F814W colour map in Fig.\,\ref{NGC1386}, middle right). The nucleus is the bright source nearly at the bottom part of this filament. The map also shows a network of interlacing filaments, all located preferentially at the Western side of the nucleus, and further extending over kpc scales (Fig.\,\ref{NGC1386}, top right panel). The morphology suggests that the filaments are not coplanar. Most of this large-scale dust structure is located at the inner rim of the circumnuclear star-forming ring at a radius of $\sim 0.5\, \rm{kpc}$ (\citealt{2000ApJS..128..139F}; Fern\'andez-Ontiveros et al. in preparation).

In previous studies, the nucleus was assumed to fall at the location of the brightest blob in \textit{HST} H$\alpha$ and [\textsc{O\,iii}] $5007\, \rm{\AA}$ line images (\citealt{2000ApJS..128..139F}), and \cite{2009ApJ...703..802M} assume the radio core to be also on this location. Fig.\,\ref{NGC1386}, bottom left, shows the correct registration of the H$\alpha$ emission relative to the location of the IR nucleus. The brightest H$\alpha$ blob is about $0\farcs2$ ($15\, \rm{pc}$) South of the IR nucleus and no H$\alpha$ counterpart to the nucleus is detected, as expected taking into account that this is an obscured nucleus up to at least $1\, \rm{\micron}$. The extinction in this filament is above $3\, \rm{mag}$. On the contrary, the peak of the [\textsc{Si\,vii}] $2.48\, \rm{\micron}$ coronal line falls on top of the nucleus (Fig.\,\ref{NGC1386} bottom right panel). While the morphology of H$\alpha$ is clearly determined by the location of the dust, that of [\textsc{Si\,vii}] extends smoothly at both sides of the nucleus, along the major axis of the galaxy. Thus the apparent collimation of the optical ionised gas in this galaxy is determined by the dust location, the AGN ionised gas may extend isotropically along the plane of the galaxy.
 

\subsection*{NGC\,3169}
NGC\,3169 is a LLAGN \citep{1995ApJS...98..477H} and as such, the galaxy light largely dominates that of the nucleus. Futhermore, the nucleus of NGC\,3169 is obscured. Even in these cases of nuclear obscuration, $2\, \rm{\micron}$ is the pivot wavelength for detecting the nucleus (see also Sombrero). NGC\,3169 is obscured in both \textit{HST}/F547M and F814W filters but it appears as a strong source from $2\, \rm{\micron}$ on and longward up to $10\, \rm{\micron}$ \citep{2012JPhCS.372a2006F}. The offset between the optical and IR peaks is $140 \pm 20\, \rm{mas}$ ($17 \pm 3\, \rm{pc}$; Fig.\,\ref{NGC3169}, middle left panel).
The colour map \textit{Ks}-band/F814W in Fig.\,\ref{NGC3169}, middle right, discloses the nuclear dust lanes presumably causing the obscuration. Specifically across the centre, it is possible to identify a relatively thin dust lane getting through the nucleus.

The ionised gas, presenting conical morphology in e.g. \textit{HST} H$\alpha$ in Fig.\,\ref{NGC3169}, bottom panel, has the peak emission next to the position of the IR photometric peak. The shift is $140 \pm 20\, \rm{mas}$ and is caused by obscuration by the dust lane (inferred $A_V \gtrsim 2.7\, \rm{mag}$, Table\,\ref{extinction}). 
The figure also shows that most of the nuclear dust is located at the Western side of the nucleus and further obscures a large fraction of the H$\alpha$ counter cone, while some H$\alpha$ further South reappears at regions where the dust is depressed (darker regions in the colour map). These central dust lanes are seen forming part of a much larger scale dust distribution detected up to several kpc from the centre (Fig.\,\ref{NGC3169}, top right), all on the Western side. Also in this galaxy, the apparent collimation of the ionised gas is produced by the relative location of the dust.

\subsection*{Sombrero galaxy (NGC\,4594)}
Sombrero is the second LLAGN in this study, along with NGC\,3169 (former sub-section, \citealt{1995ApJS...98..477H}). Contrarily to all other objects in this study, the nucleus of Sombrero is visible at all wavelengths including the UV, still it is surrounded by plenty of dust in common with all other objects. Thus, Sombrero is further used as a test of our astrometry procedure. As in all other galaxies, the nucleus, although visible, is not used for the image registration. Instead, three stellar clusters common to all images are used (Fig.\,\ref{NGC4594}, Table\,\ref{filtros}).

The nucleus is a point-like source in all the \textit{HST} and NaCo images examined, from UV (\textit{HST}/ACS\,F250W filter, \citealt{2005ApJ...625..699M}) to \textit{K}-band. The peak emission is reached, as in all other galaxies in our sample, in the $1$--$2\, \rm{\micron}$ range \citep{2012JPhCS.372a2006F}. It is however undetected beyond $10\, \rm{\micron}$ at the sensitivity limit --less than a few Jy-- of our VLT sub-arcsec observations \citep{2010MNRAS.402..879R}. At the level of uncertainty of our astrometry, the nucleus does not show a wavelength-dependent position shift larger than $10\, \rm{mas}$ (Table\,\ref{filtros}).

The dust distribution at the centre is illustrated by the colour map \textit{K}-band/F435W shown in Fig.\,\ref{NGC4594}, middle right. A network of dust filaments, extending over $400\, \rm{pc}$ in total, cross the centre and distribute along the major axis of the galaxy. The nucleus --unobscured-- shows in the map slightly bluer compared to its surroundings (see also an equivalent colour map in \citealt{2000A&A...357..111E}). This central dust does not appear connected with further dust at kpc scale (Fig.\,\ref{NGC4594}, large-scale colour map, top right). The tight confinement of the dust filaments along the major axis suggests dust is located on a thin disk on the main galaxy plane; at the centre it gets warped and its filamentary structure becomes apparent.

The ionised gas, traced by an \textit{HST} H$\alpha$ line map, bottom left in Fig.\,\ref{NGC4594}, shows a strong point-like source coincident with the position of the nucleus. Further extended H$\alpha$ distributes, as the dust, preferentially along the main plain of the galaxy (see also \citealt{2011A&A...527A..23M}). The morphology of this extended emission follows closely that of the dust as it can be seen in the figure (see also \citealt{2000ApJ...532..323P}). The estimated extinction from the colour maps caused by these filaments is very low, $\lesssim 1\, \rm{mag}$ (Table\,\ref{extinction}), consistent with no attenuation of H$\alpha$ and the nucleus.

\subsection*{NGC\,7582}
The nuclear region of NGC\,7582 is dominated by a large complex of young star clusters interlaced by dust. The nucleus of this galaxy has always been confused with one of the brightest young star clusters in the unobscured Western side of the nuclear region within the central $100\, \rm{pc}$. With high-spatial-resolution VLT/NaCo images, the nucleus is unambiguously identified with an extremely prominent $2\, \rm{\micron}$ source located at the obscured Eastern side of the nuclear region (Fig.\,\ref{NGC7582}, middle left panel; see also \citealt{2007MNRAS.374..697B}, \citealt{2010MNRAS.402..724P}). At this location, we find an optical counterpart at $0.6\, \rm{\micron}$ in the \textit{HST}/F606W image (Fig.\,\ref{NGC7582}, top and middle left panels). This is a faint source very similar to the many star clusters in the region. The comparison of two \textit{HST} images acquired in 1995 and 2001 with the same F606W filter allowed us to detect a flux increase of a factor of $\sim 1.5$ for this source. In the IR, the nucleus reaches its maximum from $2\, \rm{\micron}$ onwards up to $20\, \rm{\micron}$, the later being the wavelength limit of high-angular-resolution images available for this source \citep{2010MNRAS.402..879R}. The high-spatial resolution SED of this nucleus is presented in \cite{2010MNRAS.402..724P}.

By performing astrometry with \textit{HST} optical and NICMOS/IR images on two nuclear star clusters, \citet{2007MNRAS.374..697B} found the location of the nucleus --the IR peak-- to be located in the region of large dust concentration. We repeat here the analysis using four clusters, the same optical \textit{HST}/F606W image and our AO NaCo/IB$2.06\, \rm{\micron}$ image (Fig.\,\ref{NGC7582}, top left panel). In this case, the NaCo filter IB$2.06\, \rm{\micron}$ is an intermediate-band pure-continuum filter. We find the optical \textit{HST}/F606W and IR peaks to coincide with each other within a precision of $\sim 40\, \rm{mas}$ ($40\, \rm{pc}$, Table\,\ref{filtros}).

The distribution of dust in the galaxy is illustrated in the colour map F160W/F606W (Fig.\,\ref{NGC7582}, top right, middle right). The contrast in the NICMOS F160W image is slightly higher than in the NaCo narrow-band image IB$2.42\, \rm{\micron}$, hence the former was preferred for analyzing the dust distribution. Dust is seen everywhere in the galaxy. The central $400\, \rm{pc}$ region shows the highest optical thickness dust lanes mingled with the disk of star clusters spreading all over this region. These dust lanes are preferentially seen at the Eastern side of the galaxy, which is where the nucleus locates in projection. The preferential location of these dust lanes is possibly an orientation effect caused by the high inclination of the disk in the galaxy.

The extinction inferred from the colour map at locations next to the nucleus position is relatively high, $A_V \gtrsim 6\, \rm{mag}$ (Table\,\ref{extinction}). This high extinction is consistent with the morphology and orientation of the [\textsc{O\,iii}] emission (presented in \citealt{2009MNRAS.393..783R}): it shows a wide angle conical shape along the rims of the dust lanes at the Western side of the nucleus, and absent at the Eastern side. On the other hand, the IR high-ionisation coronal line [\textsc{Si\,vii}] shows a symmetric, uniform morphology around the centre. The [\textsc{Si\,vii}] peak coincides with the \textit{K}-band peak with a precision of $60\, \rm{mas}$\footnote{Based on the position of two clusters --the same used by \citet{2007MNRAS.374..697B}-- identified in the filters IB2.06 and IB2.48.} (Fig.\,\ref{NGC7582}, bottom panel). All together indicates a isotropic distribution of the ionised gas in this AGN, with the conical shape of [\textsc{O\,iii}] being caused by the dust distribution within the central few hundred pc rather than by a nuclear torus.

\label{lastpage}
\end{document}